\documentclass{aa}  
\usepackage{natbib,twoopt}
\usepackage[colorlinks=true, linkcolor=red, citecolor=blue, urlcolor=blue]{hyperref}
\usepackage{graphicx}
\usepackage{txfonts}
\usepackage{amsmath}
\usepackage{float}
\usepackage{tablefootnote}
\usepackage{rotating} 
\usepackage[multiple]{footmisc}
\usepackage{academicons}
\usepackage{xcolor}
\usepackage{color}
\usepackage{CJKutf8}
\usepackage{lscape}
\usepackage{placeins}

\begin{document} 
\begin{CJK*}{UTF8}{gbsn}
\title{The survey of DA double white dwarf candidates based on DESI EDR}
\titlerunning{The survey of DA double white dwarf candidates based on DESI EDR}
\author{Ziyue Jiang (蒋子悦)\inst{1,2}, 
Hailong Yuan (袁海龙)\inst{1}, 
Zhongrui Bai (白仲瑞)\inst{1}, 
Mingkuan Yang (杨明宽)\inst{1,2}, 
Xiaozhen Yang (杨肖振)\inst{1,2}, 
Qian Liu (刘倩)\inst{1,2},
Yuji He (何玉吉)\inst{1,2}, 
Ganyu Li (李甘雨)\inst{1,2}, 
Yiqiao Dong (董义乔)\inst{1}, 
Mengxin Wang (汪梦欣) \inst{1}, 
Ming Zhou (周明)\inst{1}
\and
Haotong Zhang(张昊彤)\inst{1}}
\authorrunning{Ziyue Jiang et al.}
\institute{Key Laboratory of Optical Astronomy, National Astronomical Observatories, Chinese Academy of Sciences, Beijing 100101, China \\
     \email{htzhang@bao.ac.cn,yuanhl@bao.ac.cn}
     \and
     School of Astronomy and Space Science, University of Chinese Academy of Sciences, Beijing 100049, China\\
         }
\date{Received 21 Jan, 2025; accepted 22 May, 2025}
 
\abstract 
{}
{Mergers of double white dwarfs (DWDs) are considered significant potential progenitors of type Ia supernovae (SNe Ia), which serve as “standard candles” in cosmology to measure the expansion rate of the Universe and explore the nature of dark energy. Although there is no 
direct observational evidence to definitively determine the formation pathways of SNe Ia, studying the physical properties of DWDs provides valuable insights into their evolutionary processes, interaction modes, and merger mechanisms, which are essential for understanding the explosion mechanisms of SNe Ia. This study aims to identify DWD candidates through spectroscopic radial velocity (RV) measurements and analyze their physical properties based on DESI EDR.}
{We crossmatched DESI EDR with the \emph{Gaia} EDR3 white dwarf (WD) catalog to select DA spectra. We measured the spectroscopic RV using the cross-correlation function (CCF) and assessed the significance of RV variability using a chi-squared-based variability method. Spectroscopic $T_\mathrm{eff}$ and $\log g$ were derived by fitting the hydrogen Balmer lines, with 3D convection corrections applied. Orbital periods and semi-amplitudes were obtained through a Lomb-Scargle analysis of the RV time series. We interpolated WD cooling models and applied Monte Carlo simulations to calculate masses, cooling ages, radii, and their associated uncertainties. Additionally, we analyzed their photometric and spectral energy distribution properties to derive photometric temperatures and radii, which were then compared with the corresponding spectroscopic parameters.}
{We identified 33 DA DWD candidates with significant RV variability, including 28 new discoveries. Among them, we found an extremely low-mass DWD candidate and a potential triple system. For these candidates, we measured key physical parameters including $T_\mathrm{eff}$, $\log g$, mass, and radius, and estimated the orbital periods based on the available data. Of these, 17 candidates exhibit relatively clear periodic RV variability in the current data, and we report their best-fitting periods and RV semi-amplitudes.}
{}

\maketitle

\section{Introduction}

White dwarfs (WDs) represent the final evolutionary stage of stars, typically originating from stars with initial mass less than $8 M_\odot$ \citep{2010A&ARv..18..471A}. After exhausting their nuclear fuel, the stars pass through the red giant phase, shed their outer layers with the remaining core gradually cooling and eventually evolve into WDs. Based on atmospheric composition and spectral features, WDs are primarily categorized into the hydrogen-dominated atmosphere white dwarf (DA), neutral helium (He I)-dominated atmosphere white dwarf (DB), continuous spectrum white dwarf (DC), ionized helium (He II)-dominated atmosphere white dwarf (DO), carbon-rich atmosphere white dwarf (DQ), metal-rich atmosphere white dwarf (DZ), and other less common types \citep{1994ASPC...60...64L,1999ApJS..121....1M,2004ApJ...607..426K,2019MNRAS.482.4570G}. Among these, we focused on DAs, the most common WDs \citep{1992ApJ...394..228B,2005ApJS..156...47L,2023A&A...677A.159T}, which are readily identifiable through observations and easier to model due to their high occurrence rate and well-defined hydrogen-rich spectral features.

Type Ia supernovae (SNe Ia) are renowned for their consistent luminosity and spectral features, and regarded as “standard candles,” enabling precise cosmic distance measurements. Their observations have revealed the accelerating expansion of the Universe, establishing dark energy as a cornerstone of cosmology and a key tool for studying cosmic structure and evolution \citep{1998AJ....116.1009R,1999ApJ...517..565P,2002PhLB..545...23C,2004ApJ...607..665R,2016A&A...594A..13P}. It is widely accepted that SNe Ia result from thermonuclear explosions of WDs in binary systems when their masses reach the Chandrasekhar limit \citep{1960ApJ...132..565H,2023RAA....23h2001L}. Two primary progenitor scenarios are widely considered. One is the classic single-degenerate model, in which a WD accretes material from a non-degenerate companion until its mass reaches the Chandrasekhar limit, triggering thermonuclear explosions \citep{1973ApJ...186.1007W,1984ApJ...286..644N,2004MNRAS.350.1301H}. The other is the classic double-degenerate model, whereby two WDs in a binary system experience orbital shrinkage due to gravitational wave radiation, eventually merging and surpassing the Chandrasekhar limit to trigger thermonuclear explosions \citep{1984ApJ...277..355W,1984ApJ...284..719I,2012ApJ...749L..11B,2014ARA&A..52..107M,2018MNRAS.473.5352L}. Additional pathways, such as the super-Chandrasekhar mass model, sub-Chandrasekhar mass model, and other special progenitor scenarios such as the core-degenerate model and dynamically driven double-degenerate double-detonation ($D^6$) model, have also been proposed \citep{2011MNRAS.417.1466K,2012PASA...29..447M,2012NewAR..56..122W,2018ApJ...865...15S,2023RAA....23h2001L,2024A&A...689L...6W}. However, there is still a lack of consensus on the specific properties of SNe Ia progenitors and their explosion mechanisms \citep{2013FrPhy...8..116H,2016IJMPD..2530024M,2019NewAR..8701535S,2023RAA....23h2001L}. The type Iax supernova (SN Iax) is a distinct subclass of SNe Ia, distinguished by its lower luminosity and unique spectral characteristics \citep{2013ApJ...767...57F}. Potential progenitors include single-degenerate (such as a WD+He star, linked to LP 40-365), double-degenerate, and other models, partially explaining their properties \citep{2013A&A...559A..94W,2017Sci...357..680V,2019MNRAS.489.1489R}. However, like SNe Ia, the specific progenitor systems and explosion mechanisms of SNe Iax still remain incompletely understood.

We are particularly interested in the double white dwarf (DWD) merger channel, which plays a crucial role in various areas of astronomical research. \citet{1994ApJ...420..336Y} predicted that the merger rate of DWDs with a total mass exceeding the Chandrasekhar limit matches the formation rate of SNe Ia. Moreover, DWDs are not only potential progenitors of SNe Ia but also ideal gravitational wave sources for space-based observatories\citep{2005ApJS..156...47L,Li_2020,2023ApJS..264...39R,2023MNRAS.519.2552G}. Studying DWDs can also reveal the physical mechanisms governing the evolution from common envelope phases to binary mergers\citep{1995MNRAS.272..800H,1988ApJ...329..764L,2023A&A...669A..82L}. Currently, more than 300 DWDs have been identified, accounting for less than 5\% of the over 30,000 known WDs\citep{2024MNRAS.532.2534M}. However, DWDs should account for about 10\% of the total WD population in the Milky Way, according to the incomplete statistics from \citet{1999MNRAS.307..122M,2017MNRAS.467.1414M,2020A&A...638A.131N,Korol_2022,2024MNRAS.527.8687O,2024MNRAS.532.2534M}. This significant discrepancy arises because WDs are faint and difficult to detect. Additionally, their broad spectral absorption lines require extensive high-resolution spectra to precisely measure and confirm radial velocities (RVs) and their variability. Therefore, it is challenging to find double-degenerate progenitors, especially for low-resolution spectra\citep{2019MNRAS.482.3656R}, which also indicates that numerous undiscovered DWDs await further investigation and exploration.

Spectral fitting of RV and atmospheric parameters is one of the most common methods of studying DWDs. As early as two to three decades ago, \citet{1988ApJ...334..947S}, \citet{1989LNP...328..163F}, \citet{1990ApJ...365L..13B}, \citet{1995MNRAS.275L...1M}, and \citet{2000MNRAS.319..305M} searched for DWDs and their candidates by identifying RV variability, although the sample size was limited at that time. The advent of large-scale new-generation spectroscopic surveys has yielded a wealth of high-quality WD spectra. To test the double-degenerate channel for SNe Ia formation, the ESO Supernova Ia Progenitor Survey (SPY) \citep{2001A&A...378..556K,2001AN....322..411N} initiated large-scale RV measurements of WDs from the early 21st century, leading to the discovery of over 80 DWDs to date \citep{2005A&A...440.1087N,2009A&A...505..441K,2020A&A...638A.131N}.  Extremely low-mass (ELM) WDs, with masses of less than $ 0.3 M_\odot$ \citep{1988A&A...191...57P,Li_2019,Yuan__2023}, are considered key candidates for DWDs. \citet{2010ApJ...723.1072B} launched the ELM Survey in the Sloan Digital Sky Survey (SDSS).  Together with the ELM Survey South by \citet{2020ApJ...894...53K}, these efforts have identified about 150 DWDs \citep{2020ApJ...889...49B,2022ApJ...933...94B,2023ApJ...950..141K}. \citet{2024A&A...684A.103Y} also used RV variability and identified 56 newly discovered DWD candidates based on SDSS DR14 spectra. Additionally, \citet{2024MNRAS.532.2534M} initiated the double-lined DWD (DBL) survey, targeting relatively over-luminous sources in the WD cooling sequence based on \emph{Gaia} colors, and identified 34 double-lined systems through spectral fitting. \citet{2024MNRAS.532.2534M} also created the \texttt{CloseDWDbinaries} database, which contains over 300 identified DWDs, including 77 double-lined systems. Other large-scale projects, such as the Large Sky Area Multi-Object Fiber Spectroscopy Telescope (LAMOST), also include studies on DWDs \citep{10.1093/mnras/stad3100}. 

Most DWDs exhibit weak photometric variability, making them difficult to detect by photometric surveys such as the Zwicky Transient Facility (ZTF) and the one conducted by the Transiting Exoplanet Survey Satellite (TESS) \citep{2024lsst.confE...2S}. But photometric searches have still led to some significant discoveries, particularly in detecting short-period variable sources. \citet{2010ApJ...716L.146S} reported the discovery of the first eclipsing detached DWD \textbf{NLTT 11748} through pulsation searches. \citet{2014MNRAS.438.3399B} discovered the first double-lined eclipsing DWD \textbf{CSS 41177} and provided the masses and radii of the binary system, through combined modeling of the light curves from ULTRACAM and the RV from X-Shooter. ZTF has also identified more than ten DWDs to date\citep{2020ApJ...905...32B,2022MNRAS.509.4171K,2023ApJS..264...39R}, including \textbf{ZTF J153932.16+502738.8} with an orbital period of 6.9 min and \textbf{ZTF J2243+5242} with an orbital period of 8.8 min\citep{Burdge_2019,Burdge_2020}. \citet{2023MNRAS.525.1814M} also discovered the third-closest eclipsing DWD \textbf{WDJ022558.21-692025.38} with a distance of approximately 400 pc and an orbital period of 47.19 min, based on TESS. \citet{2017MNRAS.470.1894K} predicted that the powerful photometric capabilities of Gaia and the Large Synoptic Survey Telescope (LSST) will detect hundreds to thousands of short-period eclipsing DWDs.

In addition to the two commonly used methods mentioned above, there are other approaches for searching for DWDs.\citet{2021MNRAS.506.2269E} utilized the high-precision parallax (plx) and proper motion (pm) data from \emph{Gaia} EDR3 to obtain a sample of over 1 million binary systems, which included 1565 DWD candidates. Similarly, \citet{2022MNRAS.511.5462T} identified wide-separation WD binary systems within 100 pc, also based on Gaia EDR3, comprising 155 DWD candidates, and conducted an in-depth study by population synthesis fitting. Moreover, with the future deployment of gravitational wave detectors such as the Laser Interferometer Space Antenna (LISA), more DWDs might be detected through their gravitational wave radiation, and the dual detection of electromagnetic waves and gravitational waves will also become feasible\citep{2017MNRAS.470.1894K,2018ApJ...854L...1B,2025MNRAS.536.2770J}.

In this study, we searched for DWD candidates from the Dark Energy Spectroscopic Instrument Early Data Release (DESI EDR)\citep{desi2024}, using RV variability. The structure of this paper is as follows. Section \ref{sec:2} introduces the data sources and the criteria and methods for sample selection. Section \ref{sec:3} details the methods and results of measuring RV and its variability, as well as of determining $T_\mathrm{eff}$ and $\log g$. Section \ref{sec:period} describes the calculation process for orbital parameters such as period and semi-amplitude. Section \ref{sec:ph} explores the light curves and performed spectral energy distribution (SED) fitting. Section \ref{sec:dic} further discusses the results presented earlier, including an ELM DWD candidate, five known sources, and a potential triple-star system. Finally, Section \ref{sec:con} summarizes the paper.

\section{Data} \label{sec:2}
DESI is one of the largest spectroscopic survey projects currently being undertaken\citep{Dey_2019}. While its primary mission is to explore dark energy and the structure of the Universe, the highly efficient spectroscopic data acquisition capabilities of DESI have also provided a wealth of resources for stellar research. DESI's spectra cover a wavelength range from 3600 \text{\AA} to 9800 \text{\AA} with a resolution range of 2000 to 5500.
    
In this study, we used the DESI EDR spectroscopic dataset, which covers survey validation observation data from December 2020 to June 2021, containing spectra recording of 1.8 million unique targets, including 496,128 star sources\citep{desi2024}. To obtain WD spectra from DESI EDR STAR spectra, we referred to the WD candidate catalog provided by \citet{Gentile_Fusillo_2021}, based on \emph{Gaia} EDR3. This catalog used multiple selection criteria, including absolute magnitude, color index, \emph{Gaia} quality flags, visual inspection, and probabilities of being
a WD ($P_{WD}$ \citep{2019MNRAS.482.4570G}), ultimately identifying 1,280,266 reliable WD candidates. 

By cross-matching DESI EDR STAR spectra with the \emph{Gaia} EDR3 WD candidates catalog, we obtained 812 WD targets and 7631 corresponding single-exposure spectra. Due to low signal-to-noise ratios (S/Ns), which introduce significant errors in RV measurements, we computed the median S/N for each spectrum and set a threshold of 10 to filter the data. This improved the accuracy of RV calculations, while retaining as many candidate spectra as possible. 

\begin{figure}
\centering
\includegraphics[width=\hsize]{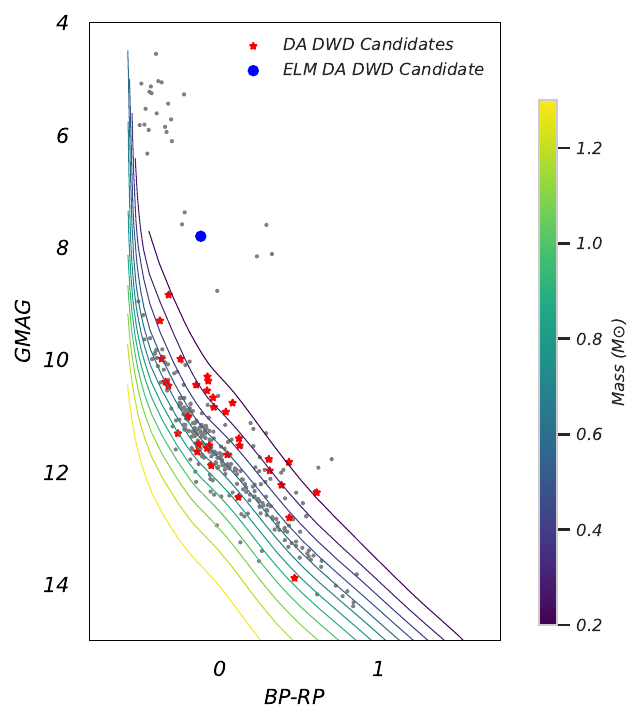}
\caption{HR diagram of 332 selected DA with S/N > 10. The red stars represent 32 DA DWD candidates, while the blue point represents the only ELM DA DWD candidate. All of these 33 sources exhibit significant RV variability. The cooling tracks were derived from the DA synthetic colors tables at \href{https://www.astro.umontreal.ca/~bergeron/CoolingModels/}{astro.umontreal.ca/\~bergeron/CoolingModels/}. Additionally, the cooling tracks are for single stars, while our candidates are binaries with a higher overall luminosity. Consequently, they fall in the region of the cooling tracks corresponding to lower masses than the ones in Section \ref{masscan}.}
\label{HR}
\end{figure}

As a result, we selected 332 DAs with 2839 spectra. The exposure time ranges from 60 s to 1800 s, with most falling between 250 s and 1400 s. The observation time span for each source ranges from half an hour to 140 days. The positions of these 332 sources and 33 DWD candidates (see Section \ref{3.4}) on the Hertzsprung-Russell (HR) diagram are shown in Figure \ref{HR}.

\section{Spectral measurements} \label{sec:3}
\subsection{Effective temperature and surface gravity}
\begin{figure}
\centering
\includegraphics[width=\hsize]{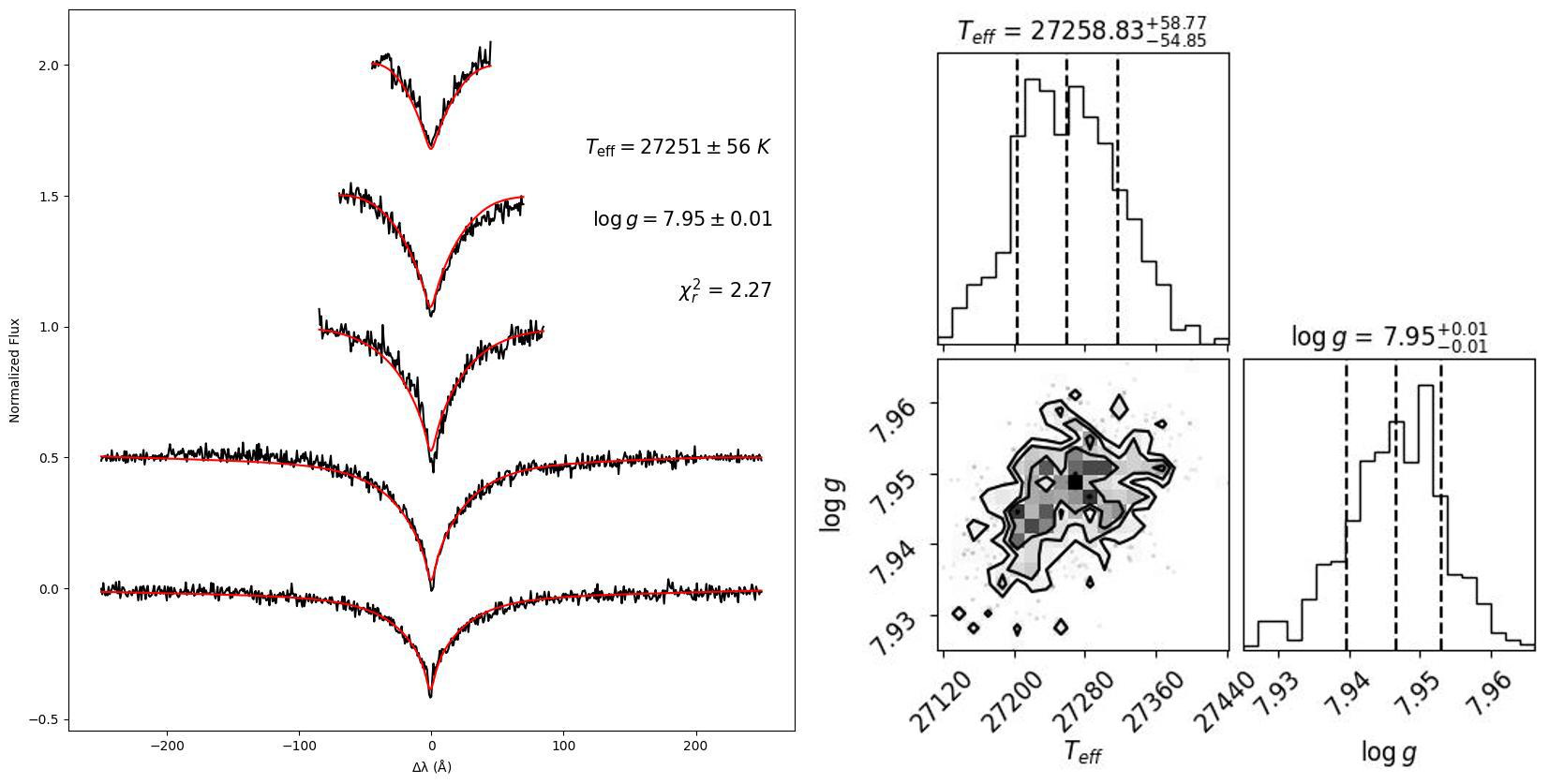}
\caption{Fitting and corner plot from \texttt{wdtools}. The left panel shows the fitting result of the hydrogen Balmer lines ($H\alpha, H\beta, H\gamma, H\delta$, and $H\epsilon$, from bottom to top) for the single-exposure spectrum of \textbf{WDJ084253.03+230025.47} at MJD = 59230.347393, yielding $T_\mathrm{eff}$ and $\log g$. The right panel is the corresponding corner plot, which illustrates the correlations between parameters and their marginal distributions, providing a visual representation of parameter uncertainties.}
\label{fit}
\end{figure}

We used the Python tool \texttt{wdtools} \citep{Chandra2020,2020zndo...3828686C} to fit the spectra of DAs in order to determine $T_\mathrm{eff}$, $\log g$, and their associated errors. The package fits the hydrogen Balmer lines (in this study, $H\alpha, H\beta, H\gamma, H\delta$, and $H\epsilon$) and interpolates the templates provided by \citet{2010MmSAI..81..921K} using a high-speed neural network. It combines the Markov chain Monte Carlo (MCMC) sampling method to derive the atmospheric parameters of WDs, effectively improving the precision of the parameter estimation. Figure \ref{fit} illustrates an example of \texttt{wdtools} results. Notably, \texttt{wdtools} shows relatively small uncertainties in $T_\mathrm{eff}$ and $\log g$ for single-exposure spectra, solely reflecting measurement errors from MCMC simulations, without implying any physical significance. When calculating the final error for an individual source, we combined the single-spectrum measurement errors and the standard deviation across multiple observations. The final error was calculated from Equation \ref{eq:4}:
\begin{equation}
\label{eq:4}
\text{error} = \sqrt{\bar{\text{er}}^2 + \text{std}^2},
\end{equation}
where $\bar{\text{er}}$ represents the mean of single-exposure spectrum measurement errors, and std denotes the standard deviation across multiple spectra.

\citet{2013AA...559A.104T} showed that 3D convective effects become particularly significant at lower temperatures (typically  $T_\mathrm{eff}$ < 15,000K), leading to a systematic overestimation of $\log g$ for cool WDs when inferred from the 1D models of \citet{2010MmSAI..81..921K}. Consequently, we implemented corrections to the spectroscopic parameters of sources in the final results with $T_\mathrm{eff}$ < 15,000K, along with the corresponding errors derived from the standard deviation of 1000 simulations. Therefore, unless otherwise specified, all spectroscopic parameters in the subsequent content have been corrected.

\subsection{RV measurement}\label{rv}
\begin{figure}
\centering
\includegraphics[width=\hsize]{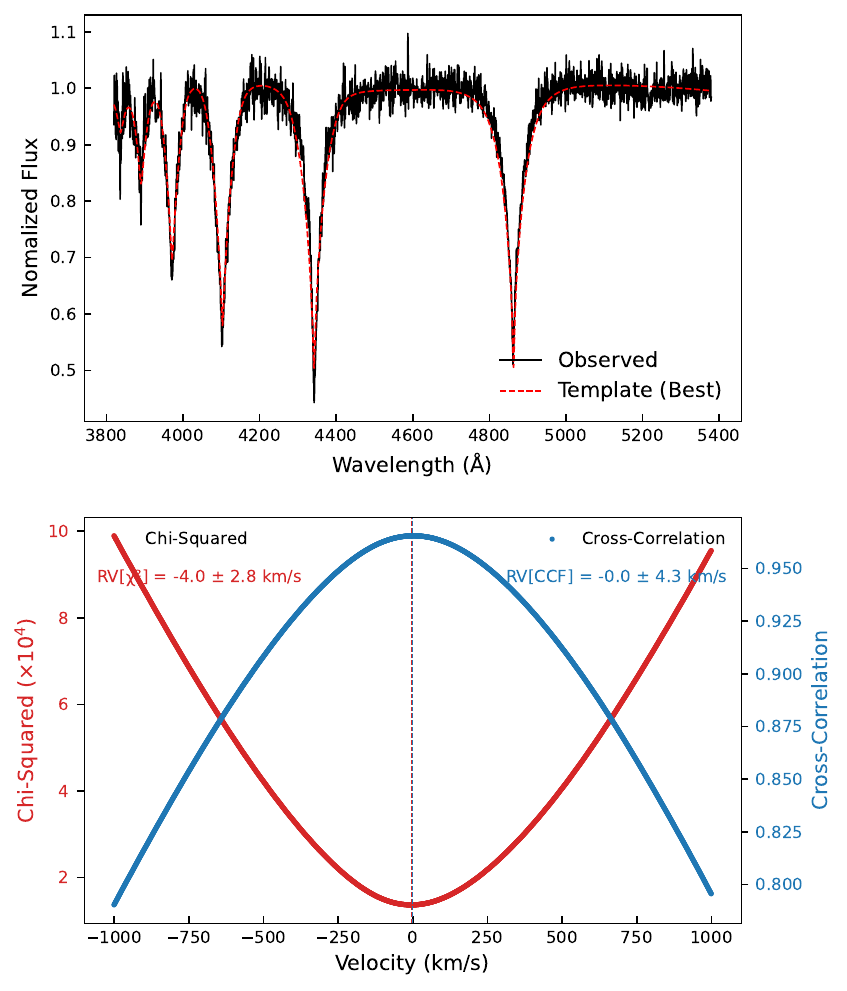}
\caption{Example of template matching and RV calculation using a single-exposure spectrum of  \textbf{WDJ084253.03+230025.47} at MJD = 59230.357356. The top panel shows the matching of the normalized spectrum with the best template ($T_\mathrm{eff} =27,000 $ and $\log g = 8.0 $ ). The bottom panel displays the $\chi^2$ curve and CCF curve obtained by shifting the best template. It can be seen that the RV results from the CCF and $\chi^2$ curves are very close.} 
\label{rvfig}
\end{figure}

We measured the RV using a template-matching method with the WD spectral templates from \citet{2010MmSAI..81..921K}, as was done in \citet{2017MNRAS.469.2102A}. We first normalized the single-exposure spectra using a polynomial fitting method. Next, we matched the normalized single-exposure spectra in the wavelength range of 3800–5500 \text{\AA} with the WD templates convolved and resampled to match the observational spectral resolution. The matching degree was evaluated using a chi-squared ($\chi^2$) function, which allowed us to determine the optimal template. Subsequently, the optimal template was shifted within a velocity range of [-1000, 1000] km/s. The most likely RV was derived from $\chi^2$ minimum and cross-correlation function (CCF) maximum \citep{Tonry1979ASO}. Figure \ref{rvfig} illustrates an example of template matching and RV calculation. The smooth $\chi^2$ and CCF curves demonstrate the quality of the fitting and matching. The final RV was taken from the velocity at CCF maximum and the velocity at $\chi^2$ minimum served as as a secondary validation.

To ensure that spectral signal errors are accurately reflected in the RV errors, we adopted the error variance formula of CCF based on a maximum likelihood approach proposed by \citet{Zucker_2003}:
\begin{equation}
\label{eq:rver}
\sigma_{s}^{2}  = -\left[ N \frac{C^{\prime \prime}(\hat{s})}{C(\hat{s})} \frac{C^{2}(\hat{s})}{1 - C^{2}(\hat{s})} \right]^{-1},
\end{equation}
where $N$ is the number of RV measurements, $C(\hat{s})$ is the value of the CCF, and $C^{\prime \prime}(\hat{s})$  is the second derivative of the CCF. Specifically, the term $\frac{C^{\prime \prime}(\hat{s})}{C(\hat{s})}$ represents the sharpness of the CCF curve peak. A larger absolute value indicates a sharper peak and a smaller error. The term $\frac{C^{2}(\hat{s})}{1 - C^{2}(\hat{s})}$ is closely related to the S/N and reflects the impact of the sample error on the velocity error. 

It is noteworthy that, due to the effects of pressure broadening and pressure shifts caused by the Stark effect, the RV measurement method proposed here is primarily suitable for subsequent variability measurements. Therefore, achieving precise RV measurements requires confirmation through high-resolution spectroscopy\citep{2020A&A...638A.131N,2024ApJ...963...17A,2025A&A...695A.131R}.

\subsection{Variability measurement}
After measuring the RV of a WD from single-exposure spectra at different time, we applied the variability detection method based on $\chi^2$ statistics proposed by \citet{2000MNRAS.319..305M} to further identify WDs with significant velocity variability. This method has been widely validated for its reliability and accuracy, particularly for analyzing velocity variability in binary systems.

For the $v_i$, the RV of the $i$-th single-exposure spectrum, we used the inverse square of its corresponding error $\frac{1}{\sigma_i^2}$, as the weight to calculate the weighted average velocity, $\bar{v}$:
\begin{equation}
\tag{3.1}
\bar{v} = \frac{\sum_{i} \frac{v_i}{\sigma_i^2}}{\sum_{i} \frac{1}{\sigma_i^2}} .
\end{equation}
Next, to quantify the deviation between $v_i$ and the weighted average velocity, $\bar{v}$, we calculated $\chi^2_{m}$:
\begin{equation}
\tag{3.2}
\chi^2_{m} = \sum_{i} \left( \frac{v_i - \bar{v}}{\sigma_i} \right)^2.
\end{equation}
We then calculated the $P$ value corresponding to the $\chi^2$ distribution with $n - 1$ degrees of freedom:
\begin{equation}
\tag{3.3}
P(\chi^2 > \chi_m^2) = 1 - F(\chi_m^2, n - 1).
\end{equation}
Here, $n$ is the number of observations and $F(\chi_m^2, n - 1)$ represents the cumulative distribution function (CDF) of the $\chi^2$ distribution. Finally, we calculated the variability index, $\eta$:
\begin{equation}
\tag{3.4}
\label{eq:eta}
\eta = -\log_{10}(P).
\end{equation}

From the above equations, it can be seen that a larger $\eta$ leads to a smaller $P$, which in turn results in a larger $\chi^2_m$, indicating a more significant velocity variability. The specific workflow and results were detailed in Section \ref{3.4}.

\subsection{Results}\label{3.4}
Considering the influence of the interval between WD template parameters, we adopted a multistep strategy involving preliminary and refined screening to minimize errors and accurately determine the RVs and their variability. 

We first performed template matching on 2839 single-exposure spectra of 332 DAs, calculating the preliminary RV and $\eta_1$ (Equation \ref{eq:eta}). Figure \ref{eta} shows the distribution histogram of $\eta_1$. Then, using the parameters from \emph{Gaia} EDR3 \citep{Gentile_Fusillo_2021}, the median values and the mean values of $T_\mathrm{eff}$ and $\log g$ derived from the single-exposure spectra of each source with \texttt{wdtools}, respectively, we fixed three templates for each source. We calculated a set of RV and $\eta$ for each template. Subsequently, we obtained three sets of RV and the corresponding $\eta_2$, $\eta_3$, and $\eta_4$. Finally, we adopted $\eta > 2.5$ (Equation \ref{eq:eta}) as the criterion for significant differences in the RV variability, taking the union of the four sets: $\eta_i > 2.5 \,(i=1,2,3,4)$.

\begin{figure}
\centering
\includegraphics[width=\hsize]{pic/eta.pdf}
\caption{Distribution histogram of $\eta_1$ values calculated from the RV of template matching. For convenience of plotting, values greater than 20 are capped at 20. In this distribution, about 13.5\% of the sources have $\eta_1 > 2.5$. Considering the theoretical prediction of DWDs accounting for around 10\% in the Milky Way \citep{1999MNRAS.307..122M,2017MNRAS.467.1414M,2020A&A...638A.131N,Korol_2022,2024MNRAS.527.8687O,2024MNRAS.532.2534M} and the need for further selection, $\eta_1 > 2.5$ is a loose but reliable criterion.}
\label{eta}
\end{figure}

After the above steps, we initially selected 63 candidate sources. For each spectrum of the 63 sources, we visually inspected the corresponding templates and the Balmer lines fit by \texttt{wdtools}, excluding the spectra with poor fits. Then, we computed the weighted average of $T_\mathrm{eff}$ and $\log g$ obtained from each single-exposure spectrum, using the S/N as the weight, to determine the atmospheric parameters for each source. Finally, we fixed the templates using the latest obtained parameters, and recalculated the RV and the $\eta_{final}$ values.

\begin{figure}
\centering
\includegraphics[width=\hsize]{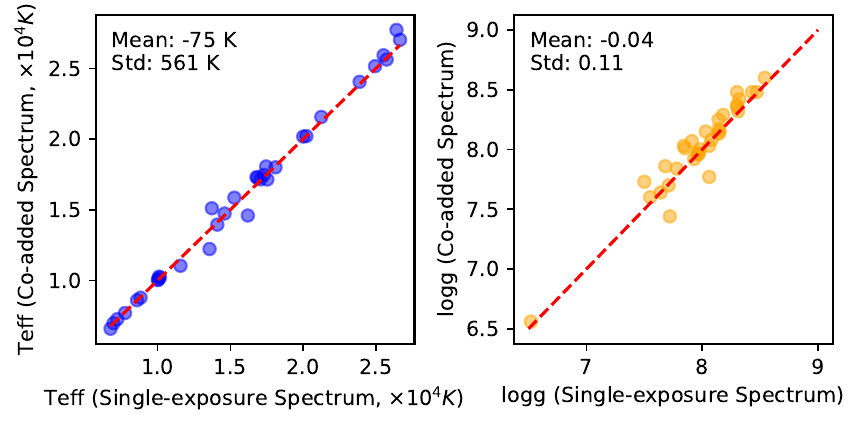}
\caption{Comparison plots between single-exposure spectra and coadded spectra. The parameters of single-exposure spectra are the weighted average of $T_\mathrm{eff}$ and $\log g$ obtained from each single-exposure spectrum, using the S/N as the weight. The parameters of coadded spectra are from \texttt{wdtools} fitting. The left panel shows $T_\mathrm{eff}$, while the right panel displays $\log g$. And $T_\mathrm{eff}$ shows a mean difference of -75 K with a standard deviation of 561 K, while $\log g$ has a mean difference of -0.04 and a standard deviation of 0.11.}
\label{com}
\end{figure}

Subsequently, we selected 33 sources with $\eta_{final} > 2.5$ as the final list of DWD candidates, as is shown in Table \ref{table:da}; their RV data are shown in Figure \ref{rv33}. Additionally, we applied a velocity correction and performed inverse-variance weighted averaging on each single-exposure spectrum to obtain the coadded spectra. We used \texttt{wdtools} to fit the coadded spectra and compared the results with the weighted results from the single-exposure spectra. Figure \ref{com} shows that the parameters remain consistent between the single-exposure spectra and the coadded spectra, as is shown in Figure \ref{coadd}.

\section{Masses and orbital parameters} \label{sec:period}
\subsection{Masses of candidates} \label{masscan}

To calculate the masses of WDs, we used \texttt{WD\_models} \citep{wdmodels} and \href{https://www.montrealwhitedwarfdatabase.org/evolution.html}{\texttt{Montreal White Dwarf Database}} \citep{2017ASPC..509....3D}. These tools both rely on synthetic color and evolutionary sequence tables of WDs, using interpolation methods to map the input values of $T_\mathrm{eff}$ and $\log g$ to parameters such as mass, cooling age, radius, and others. The former is an open-source Python package, while the latter is a convenient web application. 

Different models were applied depending on the mass range of WDs when using \texttt{WD\_models}. For low-mass models (less than about 0.5 $M_\odot$) and middle-mass models (about 0.5 to 1.0 $M_\odot$), we used the thick-H or He-atmosphere CO WD models from \citet{B_dard_2020} available at  \href{https://www.astro.umontreal.ca/~bergeron/CoolingModels/}{astro.umontreal.ca/\~bergeron/CoolingModels/} \citep{Kowalski_2006,2006AJ....132.1221H,Tremblay_2011,blouin2016newgenerationcoolwhite}. For high-mass models (greater than 1.0 $M_\odot$), we used the model for O/Ne-core WDs from \citet{2019A&A...625A..87C}. For ELMs with $\log g < 7$, we used the He-core WD model provided by \citet{2013A&A...557A..19A}.

\begin{figure}
\centering
\includegraphics[width=\hsize]{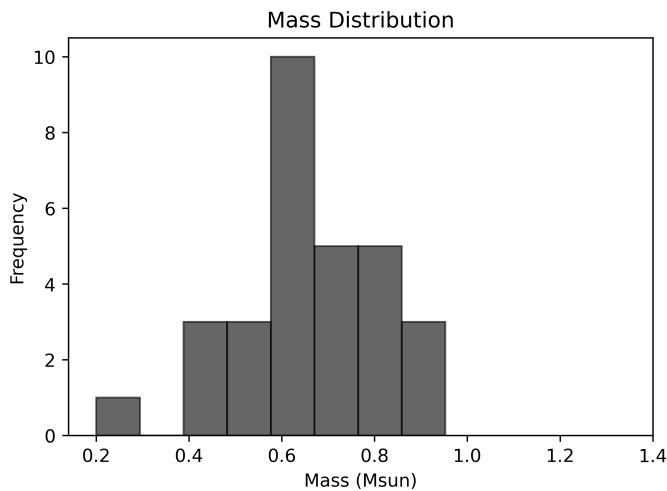}
\caption{Mass distribution histogram of DA DWD candidates. The masses of the candidates range from about $0.2 M_\odot$ to $1 M_\odot$, which is likely attributed to different evolutionary formation pathways. The peak is about $0.6 M_\odot$, which is the typical DA mass \citep{Tremblay_2011,2013ApJS..204....5K,10.1093/mnras/stv2526}.}
\label{mass}
\end{figure}

To estimate the errors of mass, cooling age, and radius, we employed the Monte Carlo sampling method. For each WD source, we assumed that the atmospheric parameters follow a normal distribution with the fit values of $T_\mathrm{eff}$ and $\log g$ as the mean and their corresponding error as the standard deviation. A Monte Carlo simulation was performed by generating 1000 random samples. Subsequently, we obtained the mass, cooling age, radius, and their respective uncertainties for 33 DA DWD candidates. Figure \ref{mass} shows the mass distribution of all candidates. There is one ELM, and no source with a mass greater than 1 $M_\odot$. Table \ref{table:da} shows a portion of the results, with more detailed data presented in \nameref{sec:data}. 

\subsection{Periods}\label{period}

In our study, we applied the Lomb-Scargle\citep{1976Ap&SS..39..447L,1982ApJ...263..835S} method to calculate the most likely periods of candidate systems. Since high-frequency (short-period) DWDs are more easily detectable and are more likely to serve as progenitors of SNe Ia \citep{2017MNRAS.468.2910B}, we focused on searching for such systems. However, when setting the frequency parameters in Lomb-Scargle method, uniform sampling in frequency results in sparse coverage of long periods, hindering the detection of long-period DWDs. Conversely, uniform sampling in periods reduces sensitivity to short-period systems. To address this issue, we developed a segmented uniform sampling strategy in periods to balance the detection efficiency between short and long periods.

We evenly distributed sampling points across three period intervals: the step size is $10^{-7}$ days for $0.02 < p \,\text{(day)} < 1$ (as the maximum spectral exposure time is 1800 s, approximately 0.02 days, making periods shorter than 0.02 days probably unreliable), $10^{-6}$ days for $1 < p \,\text{(day)} < 10$, and $10^{-5}$ days for $10 < p \,\text{(day)} < 100$. Approximately $10^7$ period points are generated in each interval and then converted to frequency points for Lomb-Scargle analysis. This distribution strategy ensures detailed coverage in the short-period range, while maintaining effective detection in the long-period range.

Due to the limited exposure times, which restricted the precision of Lomb-Scargle, we selected the five highest power points from the power spectrum as candidate periods for each source. To avoid power values becoming overly concentrated due to high-density sampling, we made the following adjustments to period selection. First, we ranked the frequencies in descending order of power and converted them to the corresponding periods. Next, to balance the distribution of high frequencies (short periods) and low frequencies (long periods), we adjusted the precision of the periods. Periods of less than 1 day were rounded to two decimal places (since periods shorter than 0.02 days are unreliable here, the minimum precision was set to 0.01 days), and periods of between 1 day and 100 days were rounded to one decimal place. Then, we removed duplicate periods, retaining only the unique period with the highest power, and recorded the period values with six decimal places of precision in the final list. Finally, we selected the five most significant periods from the power spectrum for each candidate, ensuring that they reflected the periodic features of the system. For the five candidate periods of each source, we further performed the trigonometric function fit and calculated the coefficient of determination ($R^2$). 

Ultimately, we obtained five periods and their corresponding semi-amplitudes for each candidate. We assessed the reliability of each period by visually inspecting the fitting curves and $R^2$, discarding periods with $R^2$ < 0.8. As a result, we obtained 17 DWD candidates with 1--5 well-fit periods. Detailed data and plots are presented in Table \ref{table:17} and \nameref{sec:data}. Additionally, further RVs are required to confirm these values in a robust way, due to the limited current data.

\section{Photometry} \label{sec:ph}
\subsection{The search for light curves}
We crossmatched 33 DA DWD candidates with the Zwicky Transient Facility (ZTF)\citep{2019PASP..131a8002B,2019PASP..131a8003M,2019PASP..131g8001G,10.1093/mnras/staa2814}, the All Sky Automated Survey for Supernovae (ASAS-SN)\citep{2017PASP..129j4502K}, the Transit Exoplanet Survey Satellite (TESS)\citep{2015JATIS...1a4003R}, the Catalina Real-Time Transient Survey(CRTS)\citep{2009ApJ...696..870D}, and the Wide Field Survey Telescope (WFST, “Mozi”)\citep{2023SCPMA..6609512W}. However, we did not obtain any significant light curves through Lomb-Scargle period analysis and trigonometric fitting.

\begin{figure}
\centering
\includegraphics[width=\hsize]{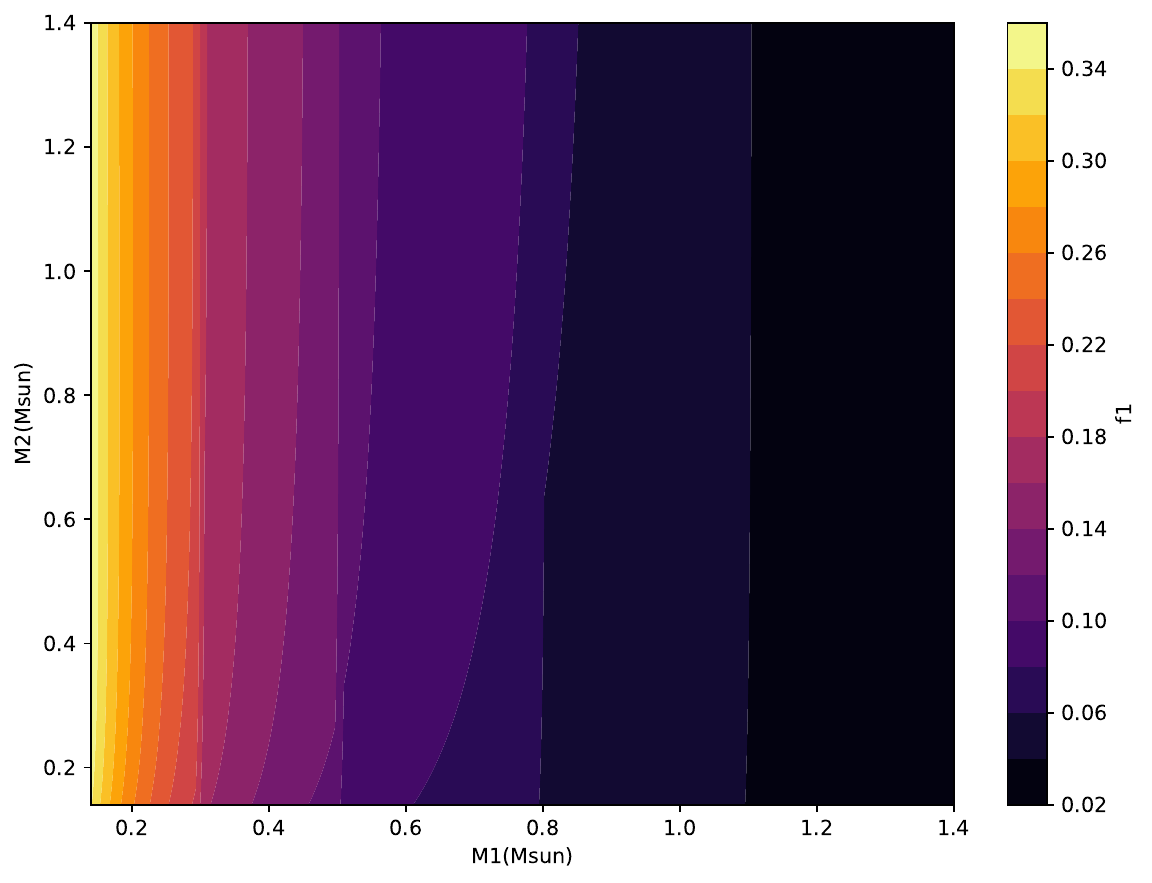}
\caption{Diagram showing how the Roche lobe filling degree of the primary star in a DWD system with a 30-minute orbital period varies with the binary mass. Here, $f_1 = \frac{R_1}{R_{L1}}$, where $R_{L1}$ is the Roche lobe radius from \citet{P.P.Eggleton1983Attr}.}
\label{lo}
\end{figure}

This is primarily because the eclipse probability of our candidates is low, and the ellipsoidal variation amplitude is small. First, most confirmed eclipsing DWD systems are compact binaries with extremely short periods. For instance, ZTF's search for DWDs primarily focuses on systems with periods shorter than one hour\citep{2022MNRAS.509.4171K}, including the previously mentioned \textbf{ZTF J153932.16+502738.8} with a period of 6.9 min and \textbf{ZTF J224342.97+524205.9} with a period of 8.8 min \citep{Burdge_2019,Burdge_2020}. Our detection has a minimum response period of approximately half an hour, so we were unable to detect candidates with periods shorter than this. Additionally, Figure \ref{lo} presented a simple simulation that we performed, revealing that in a DWD system with a 30-minute orbital period, the Roche lobe filling factor of the primary star is less than 0.36. According to \citet{2021MNRAS.501.2822G}, the ellipsoidal variation amplitude is smaller than 2\% when the filling factor is less than 0.4, which is lower than most of the errors in the above photometric surveys. Therefore, in DWD systems with periods longer than half an hour, the ellipsoidal variation amplitude should be very small.

\subsection{SEDs}
\label{sec:sed}
To further verify the accuracy of spectral fitting, we conducted the single-DA SED fitting for 32 selected DA DWD candidates. Notably, one of the 33 candidates lacked sufficient and reliable photometric data in \textbf{CDS}, as is discussed in Section \ref{3xing}. Consequently, we did not perform SED fitting for this source. We utilized a range of observational data across multiple wavelengths, including GALEX GR6+7 (filters: FUV, NUV) \citep{Bianchi_2017}, SDSS DR16 (filter: u) \citep{2020ApJS..249....3A}, Pan-STARRS1 DR1 (filters: g, r, i, z, y) \citep{2016arXiv161205560C}, and \emph{Gaia} DR3 (filters: BP, G, RP) \citep{2016A&A...595A...1G,2023A&A...674A...1G}.
In addition, for the near-infrared coverage, we adopted data from one of the following surveys depending on availability: the 2MASS All-Sky Point Source Catalog (filters: J, H, Ks) \citep{2003tmc..book.....C}, UKIDSS-DR9 (filters: Y, J, H, K) \citep{2007MNRAS.379.1599L}, or UltraVISTA DR4 (filters: Y, J, H, Ks) \citep{2012A&A...544A.156M}. The fitting was performed using the DA model spectra \citep{2010MmSAI..81..921K}. All SED fitting plots were presented in \nameref{sec:data}.

\begin{figure}
\centering
\includegraphics[width=\hsize]{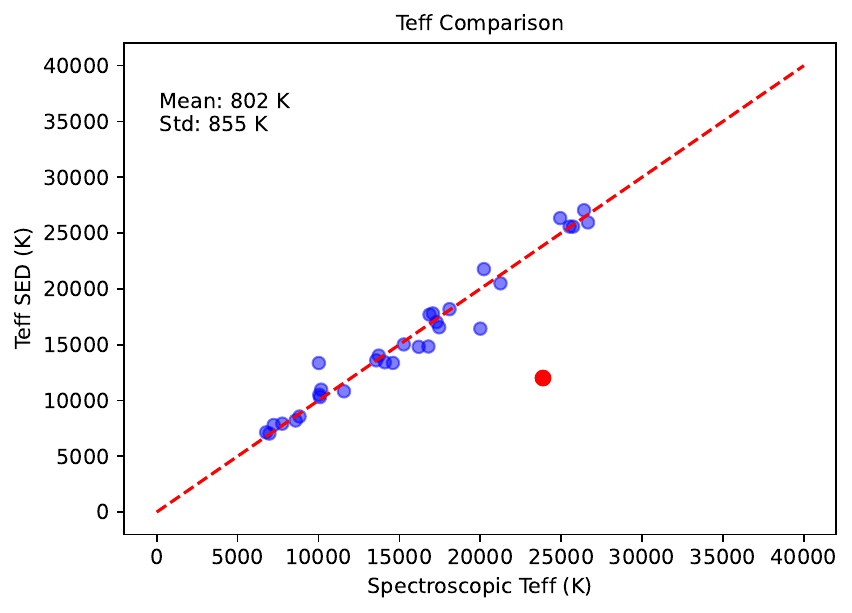}
\caption{Comparison plot between $T_\mathrm{eff}$ derived from the single-DA SED fitting and those obtained from previous spectral fitting. The red point in the plot represents the outlier \textbf{WDJ100236.84+023835.03} with significantly inconsistent temperatures, as is discussed in Section \ref{outlier}. Ignoring this outlier, the spectroscopic temperatures differ from the photometric temperatures by a mean difference of 802 K , with a standard deviation of 855 K.}
\label{comsed}
\end{figure}

To eliminate the effects of interstellar extinction, we utilized the Python package \texttt{dustmaps} \citep{2018JOSS....3..695M}, which was used to obtain the $E(B-V)$ values from the 3D \texttt{Bayestar} model \citep{2019ApJ...887...93G}. \texttt{Bayestar} was constructed using the photometric data from Pan-STARRS 1 for 800 million stars and partial stellar photometry data from 2MASS, covering a quarter of the sample \citep{2018JOSS....3..695M}. The plx values were from the \emph{Gaia} WD catalog \citep{Gentile_Fusillo_2021}. As a result, we obtained the photometric temperatures and photometric radii of these candidates. And we compared the photometric temperatures with the spectroscopic temperatures, as is shown in Figure \ref{comsed}. Except for the outlier, these two temperatures are generally consistent.

\begin{figure}
\centering
\includegraphics[width=\hsize]{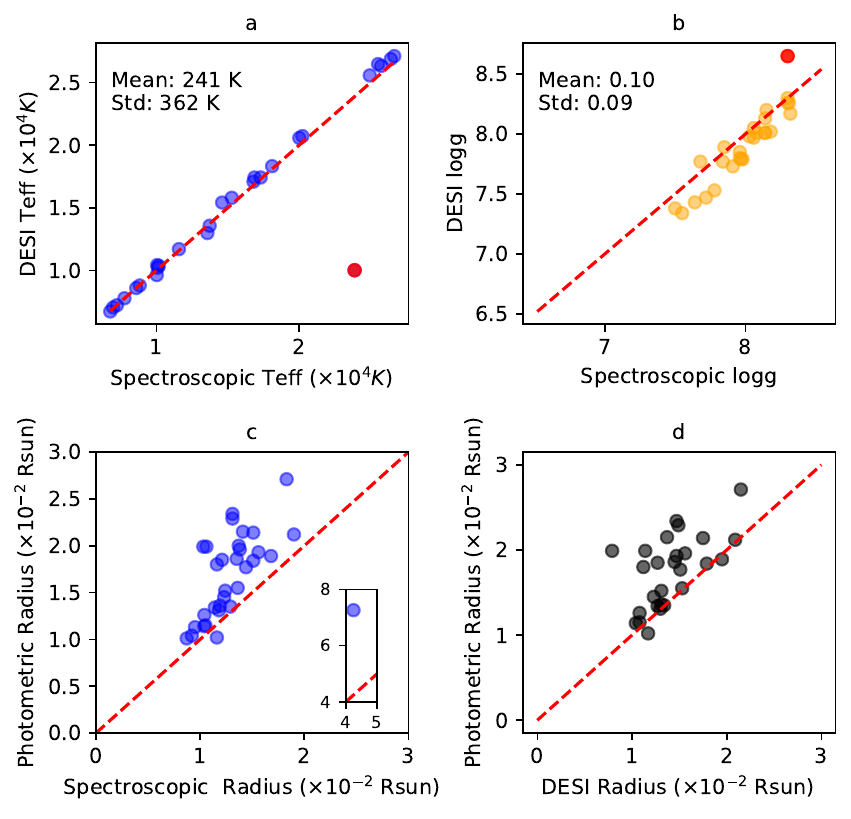}
\caption{a and b : Comparison plots between our spectroscopic parameters and those from \citet{2024MNRAS.535..254M}. The red point represents the outlier in Figure \ref{comsed}. Temporarily ignoring this outlier, our calculated $T_\mathrm{eff}$ and $\log g$ differ from their results by a mean difference of 241 K and 0.1, with a standard deviation of 362 K and 0.09, respectively, indicating a certain degree of consistency. Notably, six sources are not included in their catalog. c and d: Comparison plots of the spectroscopic radii calculated from our spectroscopic parameters and those from \citet{2024MNRAS.535..254M}, against the photometric radii. It is evident that the majority of photometric radii are larger than both sets of the spectroscopic radii.}
\label{para}
\end{figure}

Additionally, we calculated the spectroscopic radii by interpolating atmospheric parameters into the \texttt{Montreal WD Database} \citep{2017ASPC..509....3D} and compared them with the photometric radii, as is shown in panel c of Figure \ref{para}. Panel c indicates that the photometric radii we measured are significantly larger than the spectroscopic radii. To check our results, we obtained the spectroscopic parameters from \citet{2024MNRAS.535..254M}, who also conducted a study of WDs based on DESI EDR spectra. Panel a and panel b of Figure \ref{para} indicate that $T_\mathrm{eff}$ and $\log g$ from both catalogs are very close, except for the outlier. Since the value of the spectroscopic radius is highly sensitive to $\log g$, we recalculated the spectroscopic radii using their parameters and compared them with the photometric radii. Panel d confirms that the larger photometric radii were not caused by small deviations in $\log g$. These results further strengthen the case for our DWD candidates, as unresolved binary systems are often interpreted as single objects in SED fitting, leading to systematically overestimated photometric radii.

Considering the potential influence of a companion star, we conducted SED fitting using double-DA templates for these candidates. However, due to significant uncertainties in the results, we have opted not to present these findings here.

\section{Discussion} \label{sec:dic}
In this section, we discuss some special candidates in detail, including a problematic candidate, an ELM, some crossmatched ELM candidates, some known candidates, and a potential triple-star system.

\subsection{A problematic candidate}\label{outlier}
\begin{figure}
\centering
\includegraphics[width=\hsize]{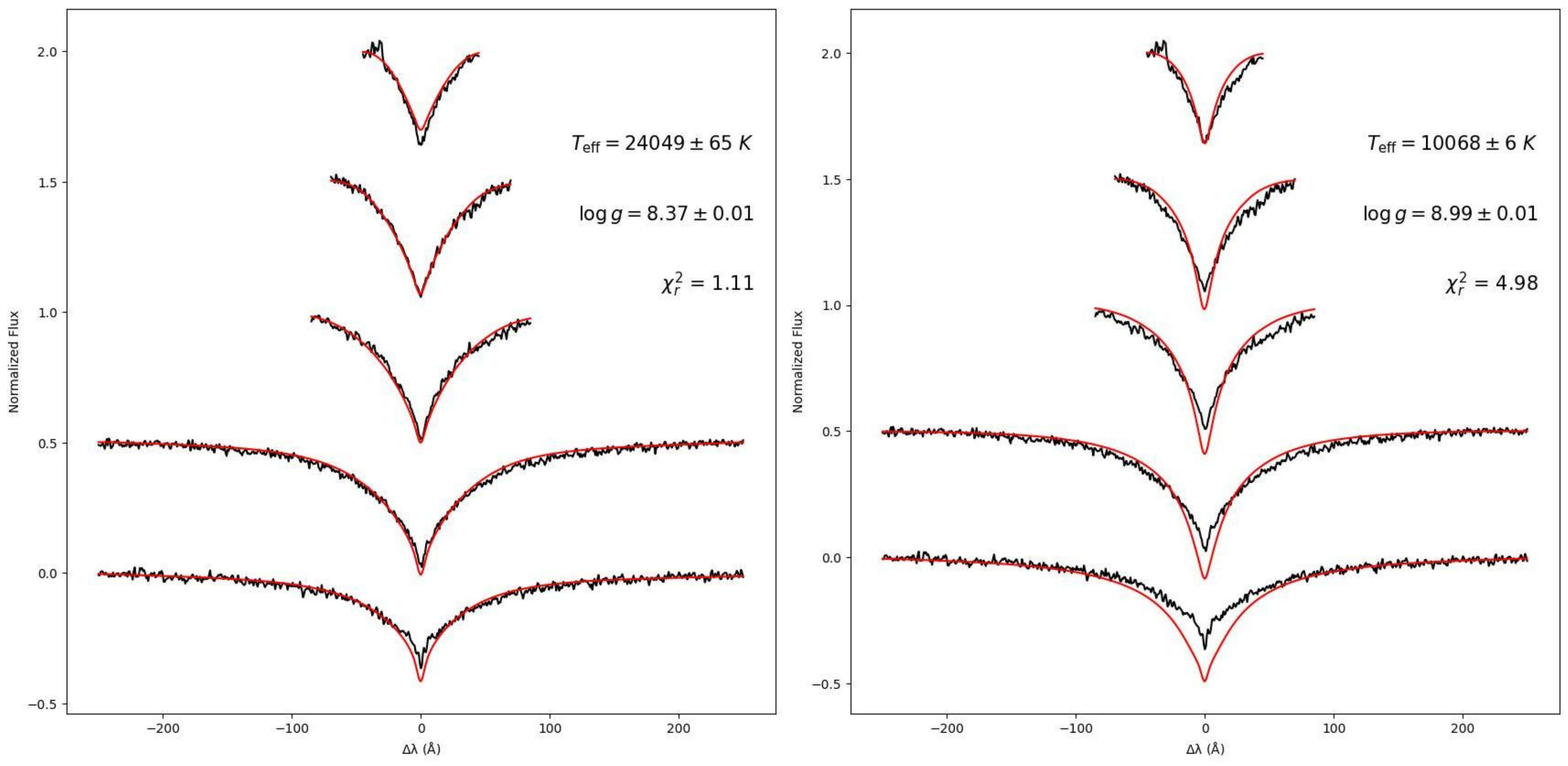}
\caption{Left panel: Best-fit result of the coadded spectrum of WDJ100236.84+023835.03 from \texttt{wdtools} without a temperature limit. Right panel: Best-fit result of the coadded spectrum of WDJ100236.84+023835.03 from \texttt{wdtools} with a temperature upper limit of 15,000 K. The results shown in the figure have not been corrected for 3D convective effects. After 3D correction, the results are 9992 K and 8.73, which is close to the results of \citet{2024MNRAS.535..254M}.}
\label{665}
\end{figure}

\begin{figure}
\centering
\includegraphics[width=\hsize]{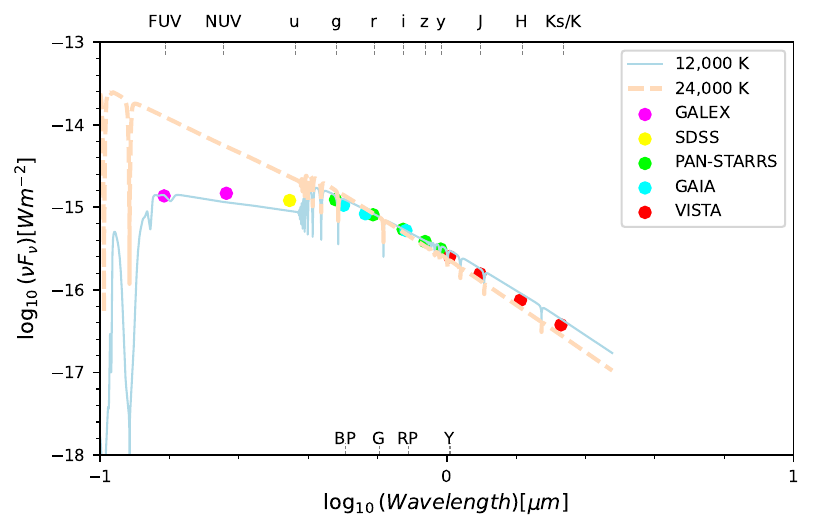}
\caption{SED fitting plot for WDJ100236.84+023835.03. The light blue curve represents the SED at about 12,000 K, while the light orange curve corresponds to about 24,000 K. In the FUV and NUV bands, the high-temperature curve is significantly higher than the observed data points.}
\label{sed}
\end{figure}

For the outlier \textbf{WDJ100236.84+023835.03} shown in Figure \ref{comsed}, we obtained its temperature of about 23,900 K, $\log g$ of 8.3, and radius of about $0.01 R_{\odot}$ from both single-exposure and coadded spectra fitting. \citet{2024MNRAS.535..254M} obtained its spectroscopic parameters with a temperature of approximately 10,008 K and $\log g$ of 8.65. When using \texttt{wdtools} for fitting with a temperature upper limit of 15,000 K, the results are 9992 K and 8.73 after 3D corrections, which are close to the ones provided by \citet{2024MNRAS.535..254M}. Figure \ref{665} shows the fitting results from \texttt{wdtools} at both high and low temperatures. Clearly, based on the spectroscopic results, the fitting at higher temperature shows a better performance. However, based on the SED results, the fitting at the lower temperature shows a better performance. From SED fitting, we obtained a temperature for it of around 12,000 K, as is shown in Figure \ref{sed} and a radius of about $0.02 R_{\odot}$.

Both spectral fitting and SED fitting match the corresponding WD template well, and the results from the double-DA model are approximately 13,000 K + 10,000 K, with the photometric radius of the primary star being about 0.015 $R_{\odot}$, which is slightly better than the single-star fit. Thus, the possibility of a binary system is a plausible explanation, although direct observational evidence is currently lacking. Therefore, the cause of such a significant temperature discrepancy between the spectroscopic and photometric results still requires further investigation using additional high-resolution observational data.

\subsection{ELM DWD candidates}

\begin{figure}
\centering
\includegraphics[width=\hsize]{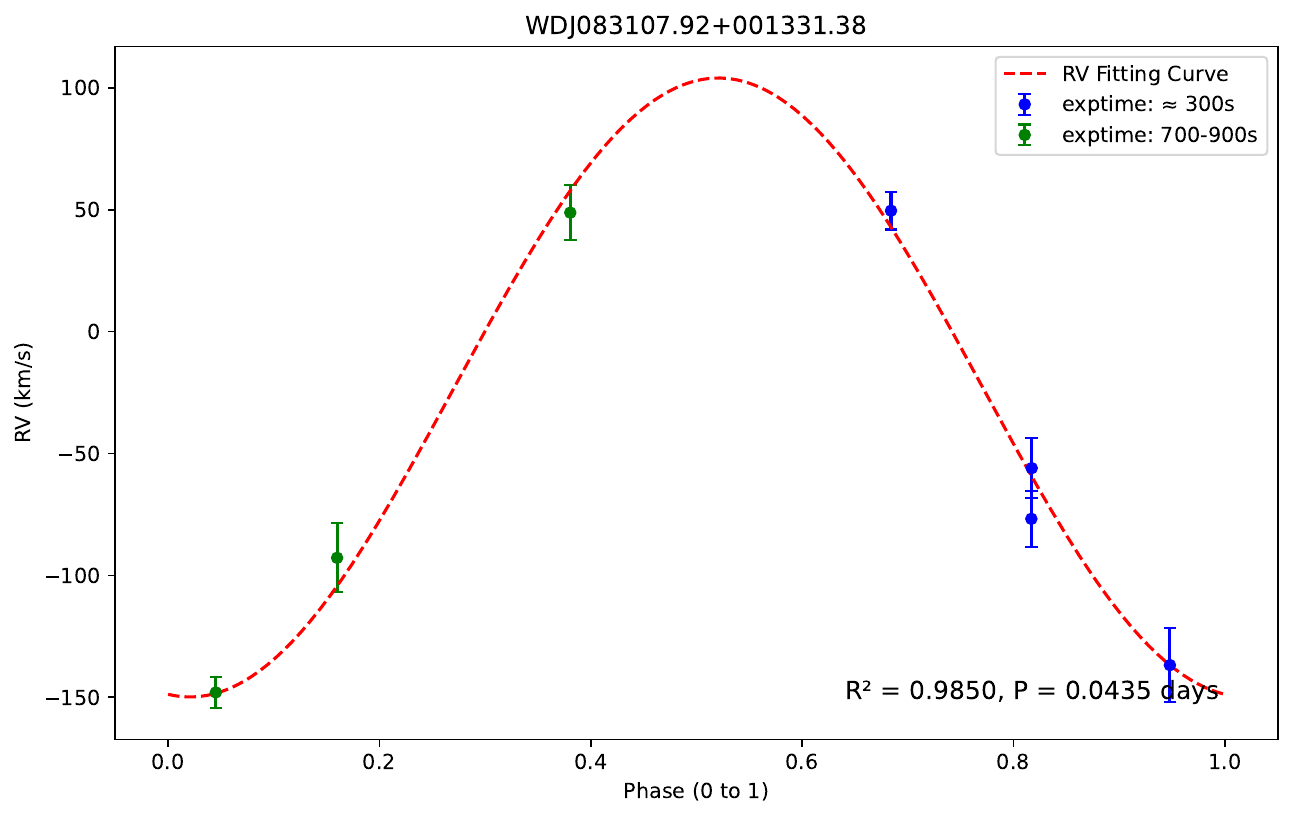}
\caption{RV fitting curve of
WDJ083107.92+001331.38. The period is 0.0435 days, approximately 3750 s. The blue points correspond to an exposure time of about 300 s, accounting for less than 1/10 of the phase, while the green points correspond to an exposure time between 700 and 900 s, covering approximately 1/4 to 1/5 of the phase.}
\label{elm}
\end{figure}

\textbf{WDJ083107.92+001331.38}, the blue point at the top of the HR diagram Figure \ref{HR}, was identified as a 0.2$M_\odot$-ELM through our multiple calculations. It exhibits significant RV variability, with a well-fit orbital period ranging from 0.0234 to 0.0532 days. Given the maximum exposure time of 900 s for its single-exposure spectra, and after examining the phase distribution and coverage, we concluded that the 0.0435-day period fit was more reliable, as is shown in Figure \ref{elm}. However, considering that there are only seven data points, additional observations are necessary to determine a more accurate period.

\begin{figure}
\centering
\includegraphics[width=\hsize]{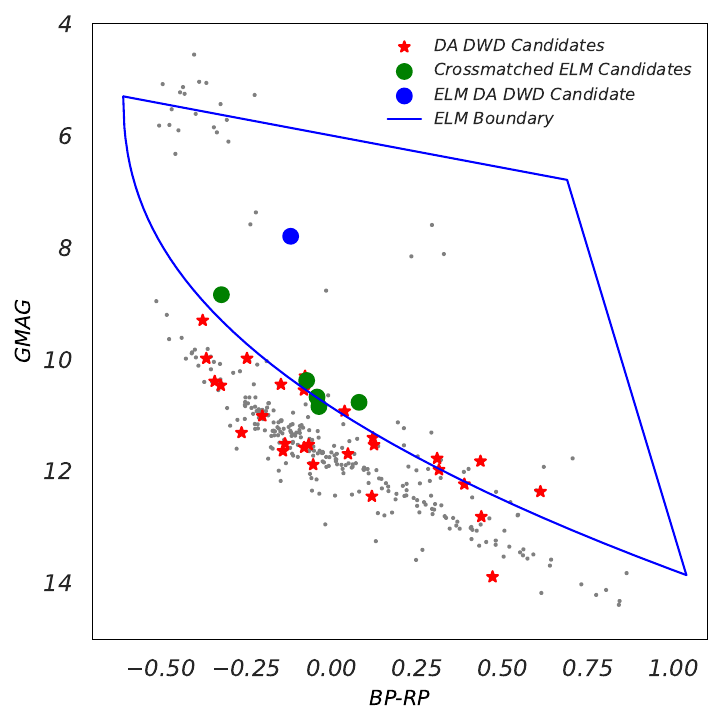}
\caption{Revised version of Figure \ref{HR}. The blue line shows the initial boundary of ELM by \citet{2019MNRAS.488.2892P}. The green points are the crossmatched ELM candidates. The blue point, WDJ083107.92+001331.38, is not present in \citet{2019MNRAS.488.2892P}, likely because it was classified as a hot subdwarf in \emph{Gaia} DR2 but as a WD in \emph{Gaia} EDR3\citep{2019A&A...621A..38G,Gentile_Fusillo_2021}.}
\label{elmcan}
\end{figure}

Moreover, to search for additional potential ELMs, we crossmatched the DWD candidates with the ELM candidate catalog based on \emph{Gaia} DR2 from \citet{2019MNRAS.488.2892P}, and identified five common sources, which are marked as green points in Figure \ref{elmcan} and listed in Table \ref{elmc}. However, the current spectroscopic $ \log g $ values of these green points are all greater than 7.0, which is beyond the typical ELM regime, as is shown in Table \ref{table:da}. Therefore, considering their $log g$ and the impact of the unseen companion, they remain potential ELM candidates rather than confirmed ELMs. Further spectroscopic observations are necessary to confirm their nature.

\begin{table}[h!]
\centering
\caption{BP-RP and GMAG of five crossmatched ELM candidates}
\label{elmc}
\begin{tabular}{c|cc}
\hline
\textbf{WDJname}& BP-RP & GMAG \\ 
\hline
WDJ092910.15+651429.78 & -0.03762 & 10.8 \\
WDJ093148.70+650304.99 & -0.04296 & 10.7 \\
WDJ093708.61+333404.69 & -0.32278 & 8.8 \\
WDJ125913.45+291653.60 & 0.07985  & 10.8 \\
WDJ141625.94+311600.55 & -0.07343 & 10.4 \\
WDJ142332.30+511156.48 & 0.30816  & 11.8 \\
\hline
\end{tabular}
\end{table}

\subsection{Known DWDs (candidates)} \label{iden}

We crossmatched our candidates with \textbf{Simbad} database and found five common sources in the literature. Among them, three have been confirmed as DWDs and two have been identified as DWD candidates.

\begin{figure}
\centering
\includegraphics[width=\hsize]{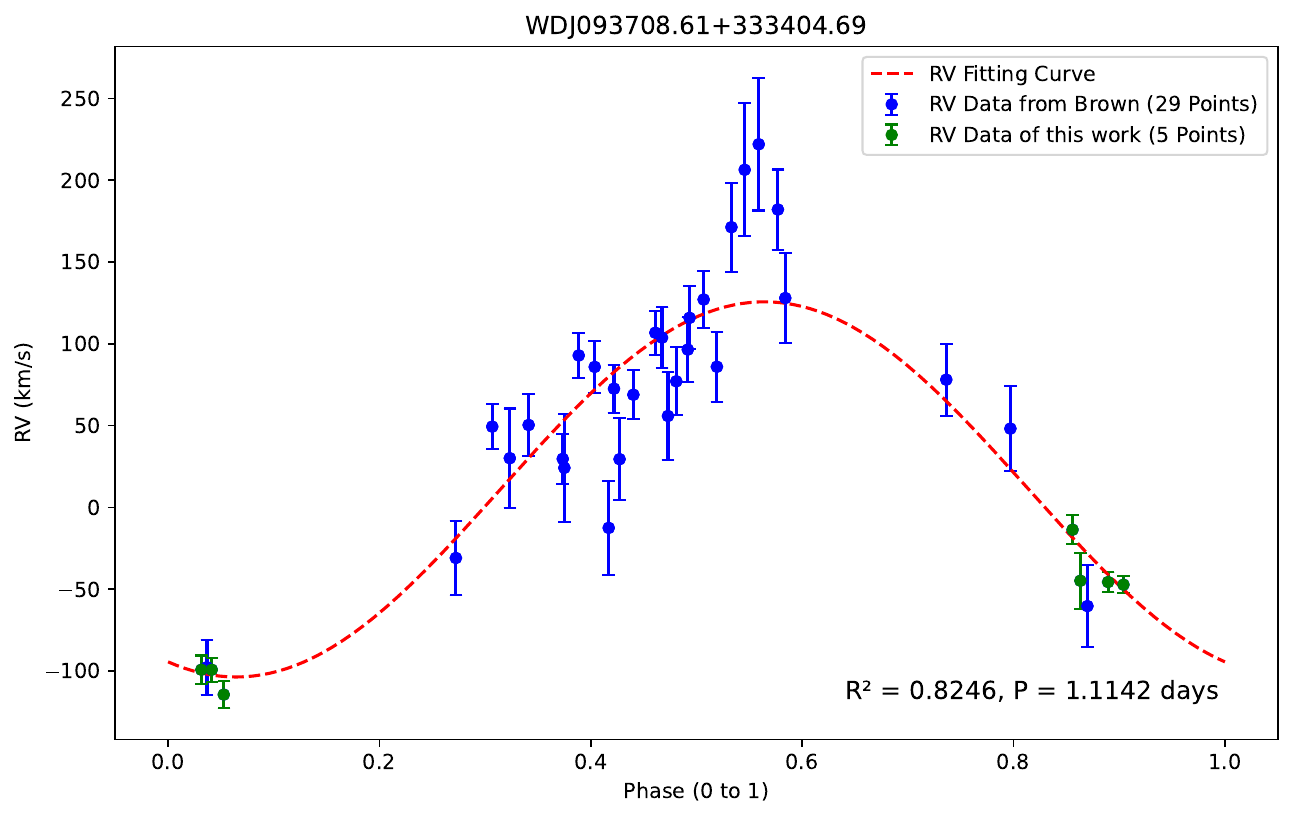}
\caption{RV fitting curve of WDJ093708.61+333404.69. The green points represent the RV data we calculated, while the blue points are from \citet{2011ApJ...730...67B}. The period is 1.114167 days, consistent with the 1.1142-day period given by \citet{2011ApJ...730...67B}, but the fit is suboptimal, with many outliers and an $R^2$ of only 0.82.}
\label{29+7}
\end{figure}

\textbf{WDJ093708.61+333404.69}: This candidate was confirmed as a DWD by \citet{2011ApJ...730...67B}, who obtained its $T_\mathrm{eff} = 24,380 \mathrm{K}$ and mass $= 0.38 M_\odot$ from \citet{2005ApJS..156...47L}, provided 29 valid RV data points, calculated the RV variability from the F-test method, and obtained a period of 1.1142 days and a minimum companion mass of 0.5 $M_\odot$. From spectroscopic measurements, we derived $T_\mathrm{eff} = 25,800 \mathrm{K}$ and $\log g = 7.55$, with a mass of $0.43 M_\odot$. Its best-fit period calculated from Lomb-Scargle is 1.0740 days. Due to only seven exposures and the limited phase coverage of the RV, the reliability of the results is constrained. We combined these 29 data points with our seven measurements and recalculated the period using Lomb-Scargle, which yielded a period of 1.114162 days, with a sinusoidal fitting $R^2$ of only 0.85, as is shown in Fig \ref{29+7}.

\begin{figure*}
\centering
\includegraphics[width=0.95\hsize]{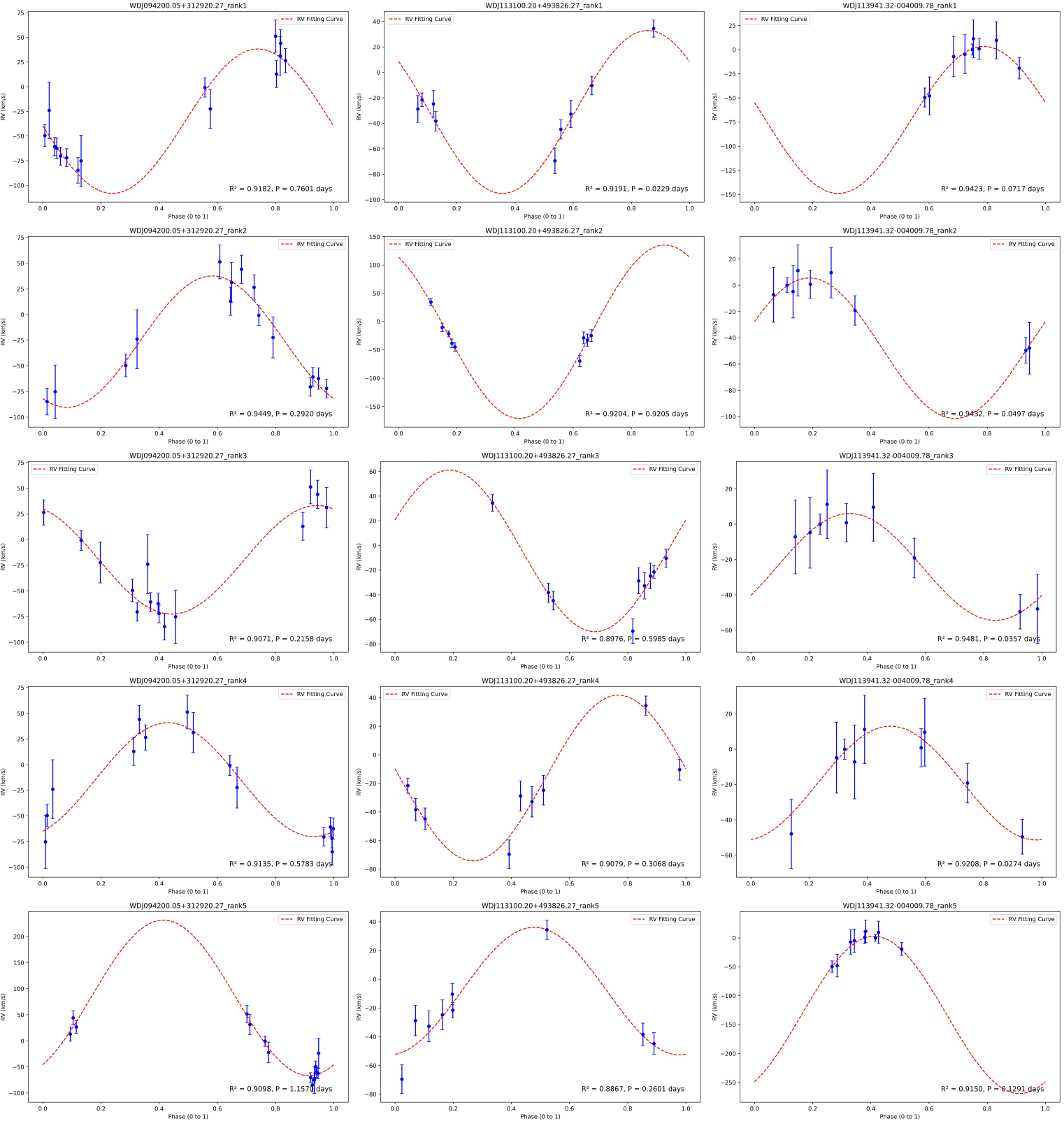}
\caption{RV fitting curves corresponding, from left to right, to WDJ094200.05+312920.27, WDJ113100.20+493826.27, and WDJ113941.32-004009.78. \textbf{WDJ094200.05+312920.27}: 15 valid exposures. The phase coverages of the rank-2 period (0.2920 days) and the rank-4 period (0.5783 days) are better compared to the other three periods. \textbf{WDJ113100.20+493826.27}: Nine valid exposures, with the best-fit period being 0.0229 days (approximately 33 minutes). However, due to the limitations of the exposure time (500s–1400s), the reliability of this period is low. Other period plots have higher $R^2$ values but suffer from low phase coverage. \textbf{WDJ113941.32-004009.78}: Nine exposures. The phase coverages of the rank-3 period (0.0357 days) and the rank-4 period (0.0274 days) are relatively good, but their accuracy is affected by the exposure time. The phase coverages of the other three periods are low. Therefore, these three DWD candidates all require additional observations to determine more accurate periods.}
\label{all}
\end{figure*}

\textbf{WDJ094200.05+312920.27}: This candidate was identified as a DWD candidate by \citet{2024A&A...684A.103Y} based on significant RV variability, but no spectroscopic parameters and orbital parameters were provided. From spectral measurements, we obtained its $T_\mathrm{eff} = 20,200\mathrm{K}$, $\log g = 7.64$, and mass $= 0.45M_\odot$. The first column of Figure \ref{all} shows its fitting curves for five periods.

\textbf{WDJ113100.20+493826.27}: This candidate was identified as a single-lined DWD by \citet{2024MNRAS.532.2534M}, who provided the parameters of the primary as $T_\mathrm{eff} \approx 10,900  \mathrm{K}$ and $\log g \approx 7.7 $, with  $T_\mathrm{eff} \approx 6900 \mathrm{K}$ of the companion. From spectral measurements, we obtained its $T_\mathrm{eff} = 10,053\mathrm{K}$, $\log g = 8.15$, and mass $= 0.69 M_\odot$. The second column of Figure \ref{all} shows its fitting curves for five periods.

\textbf{WDJ113941.32-004009.78}: This candidate was identified as a DWD candidate by \citet{2024A&A...684A.103Y} based on significant RV variability, but no spectroscopic parameters and orbital parameters were provided. From spectral measurements, we obtained its $T_\mathrm{eff} = 26,414\mathrm{K}$, $\log g = 7.5$, and mass $= 0.42M_\odot$. The third column of Figure \ref{all} shows its fitting curves for five periods.

\textbf{WDJ141625.94+311600.55}: This candidate was identified as a double-lined DWD by \citet{2024MNRAS.532.2534M}, who calculated its $T_\mathrm{eff} = 13,800 K$, $\log g = 7.86$, and mass $= 0.53 M_\odot$ with a maximum period of 1.9 days and five exposures. From spectral measurements, we obtained its $T_\mathrm{eff} = 14,602 \mathrm{K}$ and $\log g = 7.98$ with a mass of $0.60M_\odot$, also based on five exposures. Using the Lomb-Scargle method, we obtained the best-fit of 0.4127 days.

\subsection{A potential triple-star system}\label{3xing}
\begin{figure}
\centering
\includegraphics[width=\hsize]{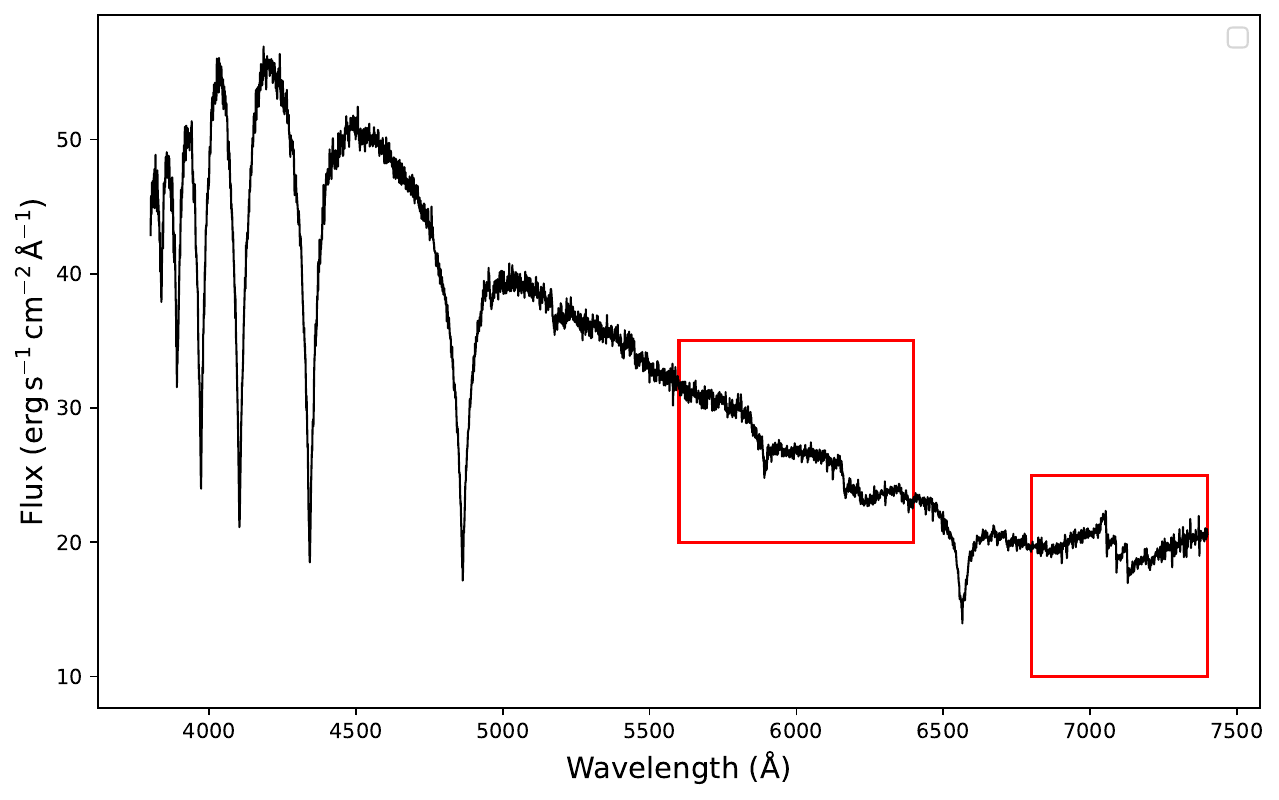}
\caption{Coadded spectrum of WDJ121654.40-012920.77. In the wavelength range above 5000 \text{\AA}, distinct features of an M-type star are observed, as is highlighted in the red boxes. This is likely caused by the contamination from the M-type star.}
\label{3m}
\end{figure}

The spectra of \textbf{WDJ121654.40-012920.77} exhibit features characteristic of an M-type star, as is shown in Figure \ref{3m}. Pan-STARRS DR1 \citep{2016arXiv161205560C} imaging also reveals a red star located about 2 arcseconds from the WD. 

From \emph{Gaia} DR3 \citep{2022yCat.1355....0G}, we obtained the BP-RP, Gmag, plx, pm, pmRA, and pmDE for both stars, as is shown in Table \ref{tab:star_params}. Additionally, from the \emph{Gaia} HR diagram, it can be inferred that this star should be an early M-type star. Based on these data, we estimate the 3D distance between the two stars to be approximately 1-5 pc. Moreover, the system satisfies the kinematic criteria of the wide binary proposed by \citet{2018MNRAS.480.4884E}, but the photometric error slightly exceeds the standard threshold and is therefore not included in the catalog of \citet{2018MNRAS.480.4884E}. Nevertheless, we still suggest that the two stars are physically associated and may form a wide binary system.

\begin{table}[h!]
\centering
\caption{Parameters of WDJ121654.40-012920.77 and the M-type Star.}
\label{tab:star_params}
\begin{tabular}{c|cc}
\hline
& WD & M-type star \\ 
\hline
BP-RP & 0.435026 & 2.435751 \\ 
Gmag & 17.9 & 15.9 \\ 
plx (mas) & 6.1391 & 6.3535 \\ 
error\_plx (mas) & 0.1317&0.0427\\
pm (mas/yr) & 41.299 & 41.698\\ 
pmRA (mas/yr) &-22.627 &  -22.43 \\ 
error\_pmRA (mas/yr) &0.173 &  0.058 \\ 
pmDE (mas/yr) & -34.549 & -35.152 \\ 
error\_pmDE (mas/yr) & 0.103 & 0.036 \\ 
\hline
\end{tabular}
\end{table}

However, at such a large 3D separation, it is implausible for the M-type star to induce the observed RV variation of $\Delta RV_{max} \approx 80 \mathrm{km/s}$. Instead, this variation is more likely caused by an unseen compact companion to the WD. Consequently, we propose that this system may be a wide triple-star system comprising a DWD and an early M-type star. Furthermore, contamination from this nearby early M-type star has compromised the photometric measurements of the WD, leaving insufficient reliable data in \textbf{CDS} for SED fitting.

\section{Conclusion} \label{sec:con}

In this study, we have searched for DA DWD candidates by measuring spectroscopic RV and selecting sources with significant RV variability. We crossmatched DESI EDR spectra with \emph{Gaia} WD catalog to select DA samples with S/N greater than 10. Significant velocity variability were identified using the $\chi^2$ variability metric, $\eta$. We fit the Balmer lines to measure the effective temperature and surface gravity, which were then interpolated to derive the masses, cooling ages, and radii. Orbital periods were calculated using the Lomb-Scargle method, with five significant periods tentatively derived for each source. These periods were further fit with trigonometric functions to derive the corresponding semi-amplitudes. We further analyzed the photometric and SED properties to complement the spectroscopic analysis and enable a direct comparison with the derived spectroscopic parameters.

Finally, we identified 33 DAs exhibiting significant RV variability as DWD candidates, of which 28 were newly discovered, including 1 ELM DA and 1 source in a potential triple-star system. For 17 of the candidates, we derived well-constrained orbital periods and semi-amplitudes based on the current data. These findings further expand the known sample of DWD candidates.

We plan to conduct follow-up observations to further study our candidates and provide valuable observational data for future SNe Ia progenitor research. We also aim to use updated data, such as DESI DR1, along with more accurate binary star models, to identify additional unknown DWDs including non-DA double degenerates in the future. 

\section*{Data availability} 
\refstepcounter{section}
\label{sec:data}
Tables in Appendix \ref{app} represent only a small subset of the full data. More complete data are available on \href{https://zenodo.org/records/15517675}{zenodo.org/records/15517675}. Appendix \ref{description} describes the online tables, which are only available in electronic form at the CDS via anonymous ftp to cdsarc.u-strasbg.fr (130.79.128.5) or via \href{http://cdsweb.u-strasbg.fr/cgi-bin/qcat?J/A+A/}{http://cdsweb.u-strasbg.fr/cgi-bin/qcat?J/A+A/}.

\begin{acknowledgements}
This research is supported by the National Natural
Science Foundation of China (12273056, 12090041,
11933004) and the National Key R\&D Program of China
(2022YFA1603002). Y.HL and B.ZR
acknowledge support from the National Key R\&D Program of China (Grant No. 2023YFA1607901).
\\

This research used data obtained with the Dark Energy Spectroscopic Instrument (DESI). DESI construction and operations is managed by the Lawrence Berkeley National Laboratory. This material is based upon work supported by the U.S. Department of Energy, Office of Science, Office of High-Energy Physics, under Contract No. DE–AC02–05CH11231, and by the National Energy Research Scientific Computing Center, a DOE Office of Science User Facility under the same contract. Additional support for DESI was provided by the U.S. National Science Foundation (NSF), Division of Astronomical Sciences under Contract No. AST-0950945 to the NSF’s National Optical-Infrared Astronomy Research Laboratory; the Science and Technology Facilities Council of the United Kingdom; the Gordon and Betty Moore Foundation; the Heising-Simons Foundation; the French Alternative Energies and Atomic Energy Commission (CEA); the National Council of Science and Technology of Mexico (CONACYT); the Ministry of Science and Innovation of Spain (MICINN), and by the DESI Member Institutions: www.desi.lbl.gov/collaborating-institutions. The DESI collaboration is honored to be permitted to conduct scientific research on Iolkam Du’ag (Kitt Peak), a mountain with particular significance to the Tohono O’odham Nation. Any opinions, findings, and conclusions or recommendations expressed in this material are those of the author(s) and do not necessarily reflect the views of the U.S. National Science Foundation, the U.S. Department of Energy, or any of the listed funding agencies. This work has made use of data from the European Space Agency (ESA) mission
{\it Gaia} (\url{https://www.cosmos.esa.int/gaia}), processed by the {\it Gaia}
Data Processing and Analysis Consortium (DPAC,
\url{https://www.cosmos.esa.int/web/gaia/dpac/consortium}). Funding for the DPAC
has been provided by national institutions, in particular the institutions
participating in the {\it Gaia} Multilateral Agreement.
\\

We acknowledge use of the VizieR catalog access
tool, operated at CDS, Strasbourg, France, and of Astropy, a communitydeveloped core Python package for
Astronomy (Astropy Collaboration, 2013). We also
acknowledge data from the Zwicky Transient Facility (ZTF, \href{https://www.ztf.caltech.edu}{https://www.ztf.caltech.edu}), funded equally
by the U.S. National Science Foundation and an international consortium of universities and institutions.

\end{acknowledgements}

\bibliographystyle{aa}   
\bibliography{references} 

\begin{appendix} 

\section{Additional material}\label{app}

\begin{figure*}
\centering
\includegraphics[width=0.95\textheight,angle=90]{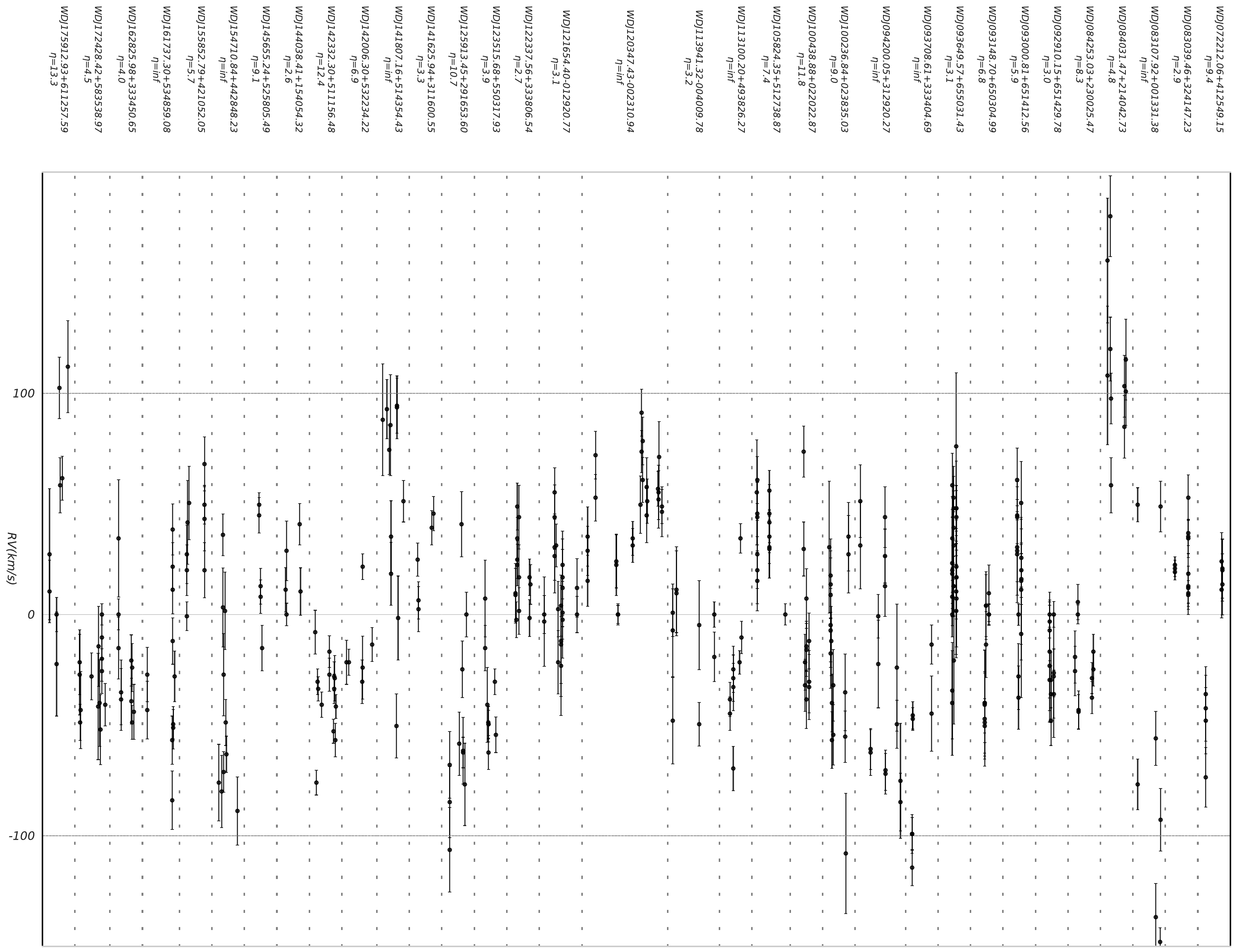}
\caption{RVs measured for the 33 DA DWD candidates with significant RV variability. The object names and the corresponding $\eta$ values are provided in the left. The "inf" means a very large $\eta$.}
\label{rv33}
\end{figure*}

\begin{figure*}
\centering
\includegraphics[width=\textwidth]{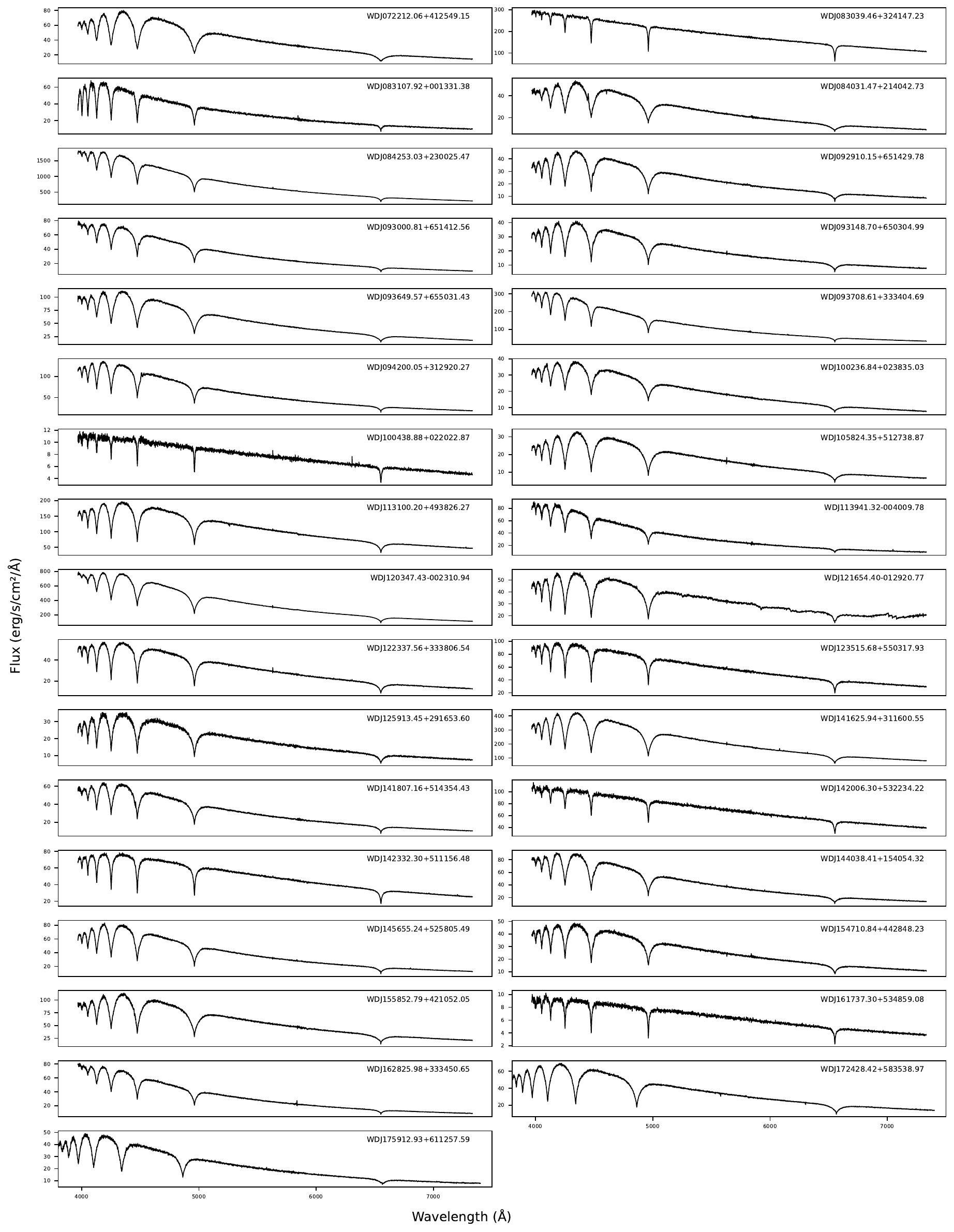}
\caption{Coadded spectra of the 33 DA DWD candidates with wavelength from 3800 \text{\AA} to 7500 \text{\AA}.}
\label{coadd}
\end{figure*}

\begin{sidewaystable}
\vspace{-11cm}
\caption{Basic parameters of 33 DA DWD candidates with significant RV variability.} 
\label{table:da} 
\centering  
\begin{tabular}{c|c|c|c|c|c|c|c|c|c}  
\hline\hline  
\textbf{WDJname} & \textbf{ra} & \textbf{dec} & \boldmath$\eta$ \tablefootnote{The "inf" in the $\eta$ column indicates that $P$ in Equation \ref{eq:eta} is very small, close to 0. When performing logarithmic calculations in Python, $\eta$ approaches infinity and is therefore marked as "inf" in our results.}& \textbf{$n_{exp}$} &\textbf{$\Delta RV_{max}(km/s)$} &$T_\mathrm{eff}(\mathrm{K})$ & $\log g$ &mass($M_\odot$)&$age_{cool}(Gyr)$\\  
\hline  
WDJ072212.06+412549.15&110.550113&41.430252&9.36&9&98&14102$\pm$647&8.43$\pm$0.13&0.88$\pm$0.08&0.513$\pm$0.139\\
WDJ083039.46+324147.23&127.663377&32.693352&2.89&12&44&6969$\pm$214&8.14$\pm$0.27&0.69$\pm$0.16&2.350$\pm$1.053\\
WDJ083107.92+001331.38&127.782978&0.225361&inf&7&198&16198$\pm$748&6.52$\pm$0.28&0.20$\pm$0.07&\textemdash\textemdash \tablefootnote{Missing parameters are due to the interpolation sequence not providing the cooling age of ELM. The parameters of this ELM were calculated using the He-core model from \citet{2013A&A...557A..19A}. If we use the CO-core model from \citet{B_dard_2020} for the calculation, the result obtained will be  mass= 0.27 $\pm$ 0.03 $M_\odot$ and $age_{cool}$= 0.013 $\pm$ 0.013 $ Gyr$. It is still an ELM.}\\
WDJ084031.47+214042.73&130.130897&21.678454&4.81&10&122&17071$\pm$713&8.54$\pm$0.08&0.95$\pm$0.05&0.354$\pm$0.064\\
WDJ084253.03+230025.47&130.720837&23.006570&8.26&10&50&26656$\pm$818&7.96$\pm$0.16&0.62$\pm$0.08&0.016$\pm$0.007\\
WDJ092910.15+651429.78&142.291777&65.241481&3.02&13&48&15274$\pm$766&7.84$\pm$0.11&0.53$\pm$0.06&0.148$\pm$0.041\\
WDJ093000.81+651412.56&142.503169&65.236727&5.93&16&98&24940$\pm$1438&8.14$\pm$0.21&0.72$\pm$0.12&0.042$\pm$0.030\\
WDJ093148.70+650304.99&142.952666&65.051292&6.81&12&60&16796$\pm$497&7.91$\pm$0.21&0.58$\pm$0.11&0.126$\pm$0.056\\
WDJ093649.57+655031.43&144.206331&65.841944&3.15&26&116&16864$\pm$781&8.30$\pm$0.17&0.81$\pm$0.11&0.255$\pm$0.088\\
WDJ093708.61+333404.69&144.285554&33.567769&inf&7&101&25737$\pm$892&7.55$\pm$0.07&0.43$\pm$0.03&0.011$\pm$0.002\\
WDJ094200.05+312920.27&145.500221&31.488876&inf&15&136&20225$\pm$477&7.64$\pm$0.11&0.45$\pm$0.05&0.032$\pm$0.007\\
WDJ100236.84+023835.03&150.653515&2.642941&8.99&17&143&23878$\pm$1071&8.30$\pm$0.17&0.81$\pm$0.11&0.076$\pm$0.040\\
WDJ100438.88+022022.87&151.161887&2.339897&11.84&11&112&6767$\pm$162&7.72$\pm$0.44&0.49$\pm$0.19&1.566$\pm$0.997\\
WDJ105824.35+512738.87&164.601249&51.460624&7.41&14&61&13569$\pm$1282&8.18$\pm$0.10&0.72$\pm$0.06&0.382$\pm$0.117\\
WDJ113100.20+493826.27&172.750593&49.640633&inf&9&104&10053$\pm$651&8.15$\pm$0.15&0.69$\pm$0.10&0.815$\pm$0.268\\
WDJ113941.32-004009.78&174.922261&-0.669505&3.17&9&61&26414$\pm$1215&7.50$\pm$0.10&0.42$\pm$0.04&0.009$\pm$0.003\\
WDJ120347.43-002310.94&180.946997&-0.386537&inf&25&91&21250$\pm$621&8.47$\pm$0.14&0.92$\pm$0.08&0.169$\pm$0.052\\
WDJ121654.40-012920.77&184.226538&-1.489306&3.14&20&78&17538$\pm$1373&7.71$\pm$0.33&0.50$\pm$0.15&0.086$\pm$0.063\\
WDJ122337.56+333806.54&185.906341&33.634979&2.68&13&51&10095$\pm$174&8.30$\pm$0.16&0.79$\pm$0.10&1.046$\pm$0.331\\
WDJ123515.68+550317.93&188.815850&55.054837&3.87&9&70&8829$\pm$64&8.08$\pm$0.15&0.65$\pm$0.09&1.016$\pm$0.235\\
WDJ125913.45+291653.60&194.805830&29.281324&10.65&10&147&10167$\pm$121&7.97$\pm$0.29&0.60$\pm$0.16&0.656$\pm$0.309\\
WDJ141625.94+311600.55&214.107822&31.266742&3.33&5&43&14602$\pm$352&7.98$\pm$0.07&0.60$\pm$0.04&0.216$\pm$0.030\\
WDJ141807.16+514354.43&214.529882&51.731748&inf&11&145&20006$\pm$827&7.96$\pm$0.11&0.60$\pm$0.06&0.063$\pm$0.021\\
WDJ142006.30+532234.22&215.026846&53.375724&6.87&6&52&7231$\pm$25&8.06$\pm$0.04&0.63$\pm$0.02&1.586$\pm$0.097\\
WDJ142332.30+511156.48&215.883157&51.198562&12.38&13&68&8589$\pm$144&7.85$\pm$0.27&0.52$\pm$0.14&0.835$\pm$0.334\\
WDJ144038.41+154054.32&220.159987&15.681602&2.61&5&41&18100$\pm$575&8.14$\pm$0.05&0.70$\pm$0.03&0.141$\pm$0.021\\
WDJ145655.24+525805.49&224.229890&52.968350&9.12&5&65&17300$\pm$258&7.78$\pm$0.05&0.50$\pm$0.02&0.077$\pm$0.009\\
WDJ154710.84+442848.23&236.795079&44.480054&inf&10&125&10023$\pm$156&8.32$\pm$0.13&0.81$\pm$0.09&1.098$\pm$0.313\\
WDJ155852.79+421052.05&239.719989&42.180941&5.65&9&69&13726$\pm$1013&8.31$\pm$0.16&0.81$\pm$0.09&0.467$\pm$0.147\\
WDJ161737.30+534859.08&244.405114&53.816475&inf&11&122&7762$\pm$92&7.68$\pm$0.34&0.46$\pm$0.16&0.957$\pm$0.497\\
WDJ162825.98+333450.65&247.108050&33.580674&3.96&10&83&25523$\pm$874&8.03$\pm$0.16&0.65$\pm$0.08&0.023$\pm$0.011\\
WDJ172428.42+583538.97&261.117660&58.594140&4.54&14&52&11576$\pm$1003&8.06$\pm$0.19&0.65$\pm$0.11&0.513$\pm$0.192\\
WDJ175912.93+611257.59&269.803801&61.215980&13.33&8&134&17460$\pm$524&7.93$\pm$0.16&0.58$\pm$0.08&0.108$\pm$0.037\\

\hline 
\end{tabular}
\end{sidewaystable}

\begin{sidewaystable}
\caption{Periods and semi-amplitudes of 17 DWD candidates with well-fit periodic curves. The unit of the period is day, and the semi-amplitude is measured in km/s.} 
\label{table:17}  
\centering  
\begin{tabular}{c|c|c|c|c|c|c|c|c|c|c}  
\hline\hline  
\textbf{WDJname} &\textbf{p1}&\textbf{k1}&\textbf{p2}&\textbf{k2}&\textbf{p3}&\textbf{k3}&\textbf{p4}&\textbf{k4}&\textbf{p5}&\textbf{k5}\\  
\hline  
WDJ083107.92+001331.38&0.025833&101.7&0.023362&102.9&0.053154&166.7&0.105125&401.9&0.04345&126.9\\
WDJ084031.47+214042.73&0.117605&67.4&0.133335&73.6&0.105195&62.1&0.153924&80.9&0.095155&58.0\\
WDJ084253.03+230025.47&0.392901&24.5&0.281881&23.7&0.648206&25.5&0.219779&23.0&1.851094&26.5\\
WDJ094200.05+312920.27&0.760139&73.2&0.291952&63.9&0.215751&52.8&0.578282&55.4&1.156983&149.0\\
WDJ100236.84+023835.03&9.985434&48.2&9.95&47.4&10.05001&49.7&9.85&45.5&10.15001&52.3\\
WDJ100438.88+022022.87&0.432368&54.8&0.759571&53.8&0.755&52.0&0.765&56.1&0.435&58.0\\
WDJ113100.20+493826.27&0.022874&64.0&0.920477&153.2&0.598473&65.5&0.306755&58.0&0.26012&44.4\\
WDJ113941.32-004009.78&0.071727&76.0&0.049662&53.5&0.035688&30.3&0.027432&32.2&0.129075&136.0\\
WDJ122337.56+333806.54&0.057603&21.6&0.054458&21.9&0.065124&20.6&0.075286&19.6&0.088667&19.2\\
WDJ123515.68+550317.93&0.094437&25.0&0.081942&26.3&0.1741&30.2&0.157703&25.1&1.137271&29.0\\
WDJ125913.45+291653.60&0.353925&74.7&0.546427&58.5&0.897612&70.8&0.177292&59.7&0.110555&60.4\\
WDJ141625.94+311600.55&0.412656&46.1&0.179616&26.1&0.048265&27.4&0.03689&67.8&0.219018&29.3\\
WDJ141807.16+514354.43&0.545691&69.6&0.250345&79.2&0.687871&87.1&0.333801&99.0&0.500696&181.3\\
WDJ142006.30+532234.22&0.177773&32.4&0.026334&26.3&0.056697&28.0&0.100007&66.0&0.068664&31.1\\
WDJ154710.84+442848.23&0.038801&67.7&0.130647&48.9&0.150352&49.0&0.115499&50.8&0.177072&50.5\\
WDJ162825.98+333450.65&0.157652&100.7&0.07883&33.4&0.085611&33.2&0.187342&91.3&0.073051&31.4\\
WDJ175912.93+611257.59&0.59808&75.5&0.374359&81.0&0.061575&61.8&1.487426&71.5&0.065616&57.2\\
\hline  
\end{tabular}
\end{sidewaystable}
\FloatBarrier

\clearpage

\section{Description of the online tables}\label{description}
\FloatBarrier
\begin{minipage}{2\columnwidth}
\textbf{Table B.1: }Catalog extension one: RVs of single-exposure spectra.
\vspace{0.2cm}
\label{fm}  
\centering 
\begin{tabular}{ccc}  
\hline
\textbf{Column}& \textbf{Heading}& \textbf{Description}\\
\hline
1&WDJname&WDJ + J2000 ra (hh mm ss.ss) + dec (dd mm ss.ss), equinox and epoch 2000\\
2&mjd& MJD identifier for a unique DESI spectrum\\
3&median\_snr& Median S/N for a unique DESI spectrum\\ 
4&velocity&RV (km/s) obtained from CCF of template-matching\\
5&velocity\_error&Error (km/s) of RV obtained from Equation \ref{eq:rver}\\
\hline
\end{tabular}
\end{minipage}

\vspace{0.5cm}
\begin{minipage}{2\columnwidth}
\textbf{Table B.2: }Catalog extension two: extended section of Table \ref{table:da}.
\vspace{0.2cm}
\label{fm}  
\small
\centering  
\begin{tabular}{ccc}  
\hline
\textbf{Column}& \textbf{Heading}& \textbf{Description}\\
\hline
1&WDJname&WDJ + J2000 ra (hh mm ss.ss) + dec (dd mm ss.ss), equinox and epoch 2000\\
2 & ra           & Right Ascension J2000 (degrees) \\
3 & dec          & Declination J2000 (degrees) \\
4 & Gmag         & \emph{Gaia} EDR3	G-band mean magnitude \\
5 & plx          & \emph{Gaia} EDR3 Absolute stellar parallax (mas) at Ep=2016.0  \\
6 & BP-RP        & \emph{Gaia} EDR3 BP-RP colour \\
7 & GMAG         & \emph{Gaia} EDR3 Absolute G magnitude \\
8 & E(B-V)       & 3D extinction obtained from \texttt{dustmaps}\citep{2018JOSS....3..695M} \\
9 & n            & Number of observations \\
10 & eta         & RV Variability parameter $\eta$ in Equation \ref{eq:eta} \\
11 & deltarv      & Maximum difference in RV \\
12 & Teff1d        & Effective temperature (K) obtained from \texttt{wdtools} \citep{Chandra2020,2020zndo...3828686C} \\
13 & Teff1d\_er      & Error of effective temperature (K) obtained from \texttt{wdtools}\\
14 & logg1d       & Surface gravity obtained from \texttt{wdtools} \\
15 & logg1d\_er      & Error of surface gravity obtained from \texttt{wdtools} \\
16 & Teff       & Effective temperature (K) after 3D corrections \citep{2013AA...559A.104T}\\
17 & logg      & Surface gravity after 3D corrections \citep{2013AA...559A.104T}\\
18 & Teff\_er       & Error (K) of Column 16\\
19 & logg\_er      & Error of Column 17\\
20 & mass        & Mass ($M_\odot$) obtained by interpolating Column 16 and Column 17 into WD evolutionary models\\
21 & mass\_er    & Error ($M_\odot$) of Column 20 \\
22 & age\_cool   & Cooling age (Gyr) obtained by interpolating Column 16 and Column 17 into WD evolutionary models\\
23 & age\_cool\_er & Error (Gyr) of Column 22 \\
24 & r\_sp       & Radius ($R_\odot$) obtained by interpolating Column 16 and Column 17 into WD evolutionary models \\
25 & r\_sp\_er & Error ($R_\odot$) of Column 24\\
26 & Teff\_sed   & Effective temperature (K) obtained from the single-DA SED fitting \\
27 & r\_sed      & Radius ($R_{\odot}$) obtained from the single-DA SED fitting \\
28 & Teff\_desi\_sp\_1d & Effective temperature (K) from DESI spectroscopic fitting by \citet{2024MNRAS.535..254M} \\
29 & logg\_desi\_sp\_1d & Surface gravity from DESI spectroscopic fitting by \citet{2024MNRAS.535..254M} \\
30 & Teff\_desi\_sp\_3d & Effective temperature (K) after 3D corrections from DESI spectroscopic fitting by \citet{2024MNRAS.535..254M} \\
31 & logg\_desi\_sp\_3d & Surface gravity after 3D corrections from DESI spectroscopic fitting by \citet{2024MNRAS.535..254M} \\
32 & r\_desi & Radius ($R_\odot$) obtained by interpolating Column 30 and Column 31 into WD evolutionary models\\
\hline
\end{tabular}
\end{minipage}

\vspace{0.5cm}
\begin{minipage}{2\columnwidth}
\textbf{Table B.3: }Catalog extension three: orbital parameters for 33 DWD candidates.
\vspace{0.2cm}
\label{fm}  
\centering  
\begin{tabular}{ccc}  
\hline
\textbf{Column}& \textbf{Heading}& \textbf{Description}\\
\hline
1&WDJname&WDJ + J2000 ra (hh mm ss.ss) + dec (dd mm ss.ss), equinox and epoch 2000\\
2&period& Period (days) obtained from the Lomb-Scargle method\\
3&rank& Power spectrum rank of the period \\ 
4&K&Semi-amplitude (km/s) corresponding to the period\\
5&R2&$R^2$ of the fitting curve.\\

\hline
\end{tabular}
\end{minipage}

\end{appendix}
\end{CJK*}
\end{document}